\documentclass[11pt]{article}
\usepackage[utf8]{inputenc}

\usepackage[margin=1in]{geometry}
\usepackage{graphicx}
\usepackage{microtype}
\usepackage{amsmath}
\usepackage[colorlinks=true, linkcolor=blue, citecolor=blue, urlcolor=blue]{hyperref}
\AtBeginDocument{%
}

\newcommand{\appref}[1]{\ref{#1}}

\usepackage[nameinlink]{cleveref}
\usepackage{float}
\usepackage{titling}
\usepackage{parskip}
\usepackage{caption}
\usepackage{subcaption}

\usepackage[backend=biber,style=authoryear,maxbibnames=2]{biblatex}
\title{\textbf{Do nineteenth-century graphics still work for today's readers?}}

\author{Yingke He}
\date{4 October 2025}

\begin{document}

\maketitle

\begin{abstract}
Do nineteenth-century graphics still work for today’s readers? To investigate this question, we conducted a controlled experiment evaluating three canonical historical visualizations—Nightingale’s polar area diagram, Playfair’s trade balance chart, and Minard’s campaign map—against modern redesigns. Fifty-four participants completed structured question-answering tasks, allowing us to measure accuracy, response time, and perceived workload (NASA--TLX). We used mixed-effects regression models to find: Nightingale’s diagram remained consistently effective across versions, achieving near-ceiling accuracy and low workload; Playfair’s dual-axis redesign underperformed relative to both its historical and alternative versions; and Minard’s map showed large accuracy gains under redesign but continued to impose high workload and long response times. These results demonstrate that some nineteenth-century designs remain effective, others degrade under certain modernizations, and some benefit from careful redesign. The findings indicate how perceptual encoding choices, task alignment, and cognitive load determine whether historical charts survive or require adaptation for contemporary use.
\end{abstract}

\section{Introduction}

Visual representations of data are essential to reasoning and decision-making, but their effectiveness depends not only on the fidelity of the data but also on the design conventions through which they are read \parencite{munzner2014visualization}. Many of today’s visualization practices trace back to canonical works of the nineteenth century, such as Florence Nightingale’s polar area diagram of Crimean War mortality, William Playfair’s trade balance charts, and Charles Minard’s flow map of Napoleon’s 1812 campaign \parencite{wood2018seeing}. These graphics are celebrated for their historical significance and are widely reproduced in teaching, public communication, and visualization research. Yet it remains unclear whether their enduring appeal reflects genuine effectiveness of design or historical prominence alone.

Understanding this question is important for both visualization research and practice. If some nineteenth-century graphics still support accurate, efficient interpretation, this suggests that they capture durable perceptual and cognitive principles. On the other hand, if others impose unnecessary difficulty relative to modern alternatives, a redesign may be essential to make them fit contemporary analytic needs. In either case, studying historical graphics offers a unique capability test for visualization design: they are familiar enough to interpret, yet distant enough in convention to reveal where usability depends on cultural or perceptual alignment \parencite{tufte1983visual}.

We conducted a controlled experiment comparing historical graphics with modern redesigns. Participants completed structured question-answering tasks based on Nightingale, Playfair, and Minard, with performance evaluated through accuracy, response time, and perceived workload. By combining objective measures with subjective assessments, and applying mixed-effects regression models, we provide a systematic account of how these classic designs perform relative to modern alternatives. The results offer insight into when historical charts continue to “work”, when redesigns confer advantages, and what these cases reveal about the broader principles of visualization design.

\section{Related Work}

Foundational research in graphical perception established that some visual encodings are decoded more accurately than others, with judgments based on position and length outperforming those based on angle or area \parencite{cleveland1984graphical}. This principle underlies many modern redesigns and helps explain why certain historical forms succeed or fail, such as Nightingale’s reliance on angular judgments compared to position-based bar charts. Building on these findings, Mackinlay formalized a ranking of encodings and showed how designs could be recommended algorithmically from data and task specifications, anticipating contemporary best-practice guidelines \parencite{mackinlay1986automating}. Subsequent large-scale web studies extended these perceptual results using crowd-sourcing, demonstrating consistent sensitivity to encoding differences across task types and noise conditions \parencite{heer2010crowdsourcing}.

Beyond low-level perception, readers’ visualization literacy also shapes performance. Instruments such as the VLAT reveal wide variability among non-experts, motivating the use of demographic adjustments and practice tasks in experimental studies \parencite{lee2017vlat}. Related work on embellishment further complicates minimalist assumptions: Previous work finds that decorated charts can match similar accuracy to plain designs while improving long-term recall, findings directly relevant both to ornate historical figures such as Minard and to modern redesigns that trade decoration for clarity \parencite{Bateman2010UsefulJunk}.

Other research has documented how misleading or complex designs can increase error. Truncated scales, parallel axes, and other compound designs have been shown to reduce accuracy, providing clear cautions for dual-axis and similar arrangements \parencite{pandey2015deceptive}. Such findings parallel the underperformance of Playfair’s modern dual-axis chart in our study.

A complementary perspective considers subjective workload alongside accuracy and time. The NASA-TLX instrument has become widely used in visualization and human-computer interaction (HCI) research to compare task structures and interface designs \parencite{hart2006nasa}. Its application links perceptual difficulty with observed performance outcomes, motivating its inclusion in our evaluation.

Although Nightingale, Playfair, and Minard are frequently cited in visualization pedagogy and public communication, most prior references remain descriptive or historical rather than empirical. Existing reproductions and commentaries emphasize narrative and context but rarely provide controlled comparisons. Direct evaluations of the original nineteenth-century forms against principled modern redesigns remain scarce.

Taken together, prior work explains why some designs should outperform others, highlights the moderating roles of reader ability and embellishment, warns about high-risk designs, and motivates workload as an outcome. What has been missing is a controlled, within-topic evaluation of canonical nineteenth-century graphics against modern redesigns. Our study addresses this gap by testing three iconic cases—Nightingale, Playfair, and Minard—under task-matched conditions, quantifying when historical forms still work and when redesigns confer measurable benefits.

\section{Methods}

\begin{figure}[H]
  \centering
  \includegraphics[width=\textwidth]{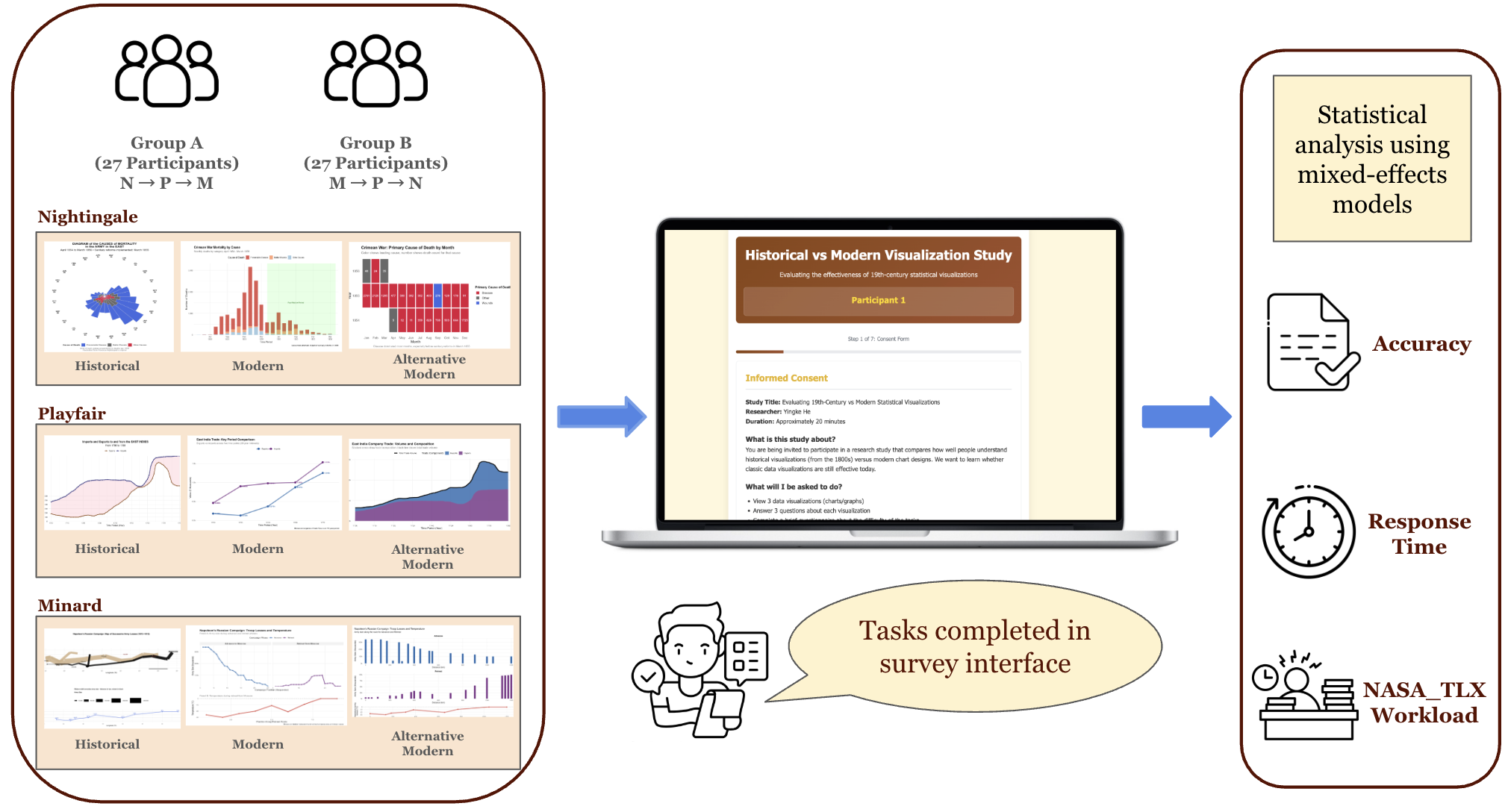}
  \caption{Study workflow overview. Participants were assigned to Group A (Nightingale→Playfair→Minard) or Group B (Minard→Playfair→Nightingale). Within each group of 27, visualization versions (historical, modern, alternative modern) were counterbalanced so that every participant received a unique combination, covering all 3×3×3 possibilities. For each dataset, participants answered three multiple-choice questions followed by a NASA–TLX workload rating. Responses were analyzed using mixed-effects regression models of accuracy, response time, and workload.}
  \label{fig:workflow}
  
\end{figure}

A controlled experiment was conducted to compare the effectiveness of three classic nineteenth-century visualizations—Nightingale, Playfair, and Minard—with modern redesigns. The study employed a mixed design that combined within-subject comparisons across datasets with between-subject variation in visualization versions. Participants completed structured question-answering tasks, and their accuracy, response times, and subjective workload were recorded and analyzed. The following subsections describe the materials, study design, participants, and analysis in detail.

\subsection{Materials}
The study materials consisted of three datasets commonly associated with classic nineteenth-century visualizations: (1) mortality causes during the Crimean War (Nightingale), (2) England’s trade balance with the East India Company (Playfair), and (3) troop losses in Napoleon’s Russian campaign (Minard). Each dataset was visualized in three versions: a recreation of the original historical design, a modern redesign, and an alternative modern redesign. 

The redesigns were not intended as independent objects of study, but rather as principled contrasts that situate the historical charts relative to contemporary design conventions. To achieve this, they were grounded in established findings from graphical perception and visualization practice \parencite{vanderplas2020goodgraph}. For Nightingale, the polar-area diagram was contrasted with a stacked bar chart and a heatmap, substituting position and color for radial area. For Playfair, the two-line chart was contrasted with a dual-axis line chart, which remains common in practice despite perceptual drawbacks, and a stacked area chart emphasizing overall trade volume and composition on a cumulative scale. For Minard, the flow map’s width-along-path encoding was replaced with small-multiples line charts and binned bar charts, both using position or length on a common scale. Figures for each version are provided in \appref{app:vis}, and screenshots of the survey interface and administration dashboard are included in \appref{app:survey} for reference.

\subsection{Study Design and Participants}

A total of 54 participants were recruited for the study. The first 27 participants were assigned to Group~A, in which visualizations were always presented in the fixed order Nightingale $\rightarrow$ Playfair $\rightarrow$ Minard. Preliminary results from this group suggested that Minard’s visualization produced very low accuracy, while Nightingale’s performed relatively well. To test whether this pattern reflected an inherent design difference or an order effect, an additional 27 participants were recruited as Group~B. In this group, the order was reversed to begin with Minard (Minard $\rightarrow$ Playfair $\rightarrow$ Nightingale).

Playfair was held constant in the middle position for both groups. This choice reflected pilot findings that Playfair’s performance was intermediate between Nightingale and Minard, and Group~B was added after the initial results as a supplementary condition to check whether Minard’s lower accuracy could be attributed to order effects rather than inherent design differences. By fixing Playfair as a stable middle condition and varying only the order of Nightingale and Minard, the design enabled a targeted and efficient check of potential sequence effects without requiring a full counterbalancing across all six possible dataset orders.

The experiment employed a mixed design. Within-subjects, each participant completed tasks on all three datasets; between-subjects, each participant was assigned to view exactly one version of each dataset (historical, modern, or alternative modern). Version assignment was counterbalanced across participants using a slot system that covered all $3 \times 3 \times 3$ combinations of dataset and version. This ensured that across each group of 27 participants, every visualization version was shown an equal number of times and that each dataset was seen by nine participants per version, preventing bias from unequal exposure.

Each dataset was paired with three multiple-choice questions of increasing difficulty (easy, medium, hard). All questions had four answer options, and responses were recorded for both correctness and response time. After completing each dataset block, participants also provided workload ratings using the NASA-TLX instrument. Detailed question design principles are provided in \appref{app:questions}.

\subsection{Measures}

Task performance was evaluated using accuracy (proportion of correct responses) 
and response time (latency from question onset to submission). These metrics 
capture both correctness and efficiency of task completion.

Following each dataset block, participants completed workload ratings using the NASA Task Load Index (NASA-TLX). Four subscales were included, each on a 10-point scale (higher = more): Mental Demand—how much thinking, remembering, and concentration the task required; Temporal Demand—how rushed the task felt and the degree of time pressure; Effort—how hard the participant had to work to achieve their performance; and Frustration—the extent of irritation, stress, or discouragement experienced during the task. NASA-TLX is widely used in HCI as a validated tool for subjective workload assessment \parencite{hart2006nasa}. 

By combining objective performance and subjective workload ratings, we obtained 
a comprehensive evaluation of visualization effectiveness.

\subsection{Statistical Analysis}

Let $i$ index trials (questions) and $j$ index participants.  
All analyses were conducted using mixed-effects regression models to account for the repeated-measures structure of the data \parencite{bates2015lme4}. The primary outcomes of interest were \textit{accuracy}, \textit{response time}, and \textit{subjective workload} (NASA-TLX).

\subsubsection*{Accuracy Model}
Accuracy was modeled as a binary variable:
\[
Y^{(\text{acc})}_{ij} \sim \text{Bernoulli}(p_{ij}), 
\quad 
\text{logit}(p_{ij}) 
= \beta_0 
+ \beta_V^\top X_{V,ij} 
+ \beta_F^\top X_{F,ij} 
+ \beta_{VF}^\top (X_{V,ij} \otimes X_{F,ij}) 
+ u_{0j},
\]
where $Y_{ij}^{(\text{acc})}$ indicates whether participant $j$ answered question $i$ correctly.  $X_{V,ij}$ encodes visualization version (historical, modern, alt-modern),  
$X_{F,ij}$ encodes dataset (Nightingale, Playfair, Minard),  
and $u_{0j} \sim \mathcal{N}(0, \sigma_u^2)$ is a participant-specific random intercept.  
Fixed effects estimate differences in log-odds of correctness across versions and datasets; exponentiated coefficients are reported as odds ratios (OR $= e^{\beta}$) with 95\% confidence intervals.

\subsubsection*{Response Time Model}
Response time $T_{ij}$ (in seconds) was log-transformed to reduce skew:
\[
\log(T_{ij}) = \beta_0 
+ \beta_V^\top X_{V,ij} 
+ \beta_F^\top X_{F,ij} 
+ \beta_{VF}^\top (X_{V,ij} \otimes X_{F,ij}) 
+ u_{0j} + \epsilon_{ij},
\]
with $\epsilon_{ij} \sim \mathcal{N}(0, \sigma^2)$.  

where $T_{ij}$ is the time for participant $j$ to answer question $i$.  The fixed and random effects structure mirrored the accuracy model. Exponentiated coefficients are interpreted as \emph{time ratios} (TR $= e^{\beta}$), e.g.\ $\exp(\beta)=1.12$ indicates 12\% longer response times.

\subsubsection*{Workload Model}
Subjective workload (NASA-TLX score, from 1-10) was modeled at the dataset level:
\[
W_{j} = \beta_0 
+ \beta_V^\top X_{V,j} 
+ \beta_F^\top X_{F,j} 
+ u_{0j} + \epsilon_j,
\]
where $W_j$ denotes average workload for participant $j$ after completing a dataset block.  Version and dataset were included as fixed effects, and participant was modeled as a random intercept.
Residuals $\epsilon_j \sim \mathcal{N}(0, \sigma^2)$, with $u_{0j}$ again capturing participant heterogeneity.

\subsubsection*{Order Effects Models}

To assess whether presentation order moderated version effects, the primary models were extended with a between-subjects order factor (Group~A: Nightingale$\to$Playfair$\to$Minard; Group~B: Minard$\to$Playfair$\to$Nightingale) and its interaction with version.

\textbf{Accuracy}:
\[
\text{logit}(p_{ij}) = \beta_0 + \beta_V X_{V,ij} + \beta_F X_{F,ij} + \beta_G X_{G,j} 
+ \beta_{VF}(X_{V,ij}\times X_{F,ij})
+ \beta_{VG}(X_{V,ij}\times X_{G,j}) + u_{0j},
\]
where $u_{0j}\sim \mathcal{N}(0,\sigma^2_u)$. The interaction term $\beta_{VG}$ captures whether version effects differ across order groups.

\textbf{Response Time}:
\[
\log(T_{ij}) = \beta_0 + \beta_V X_{V,ij} + \beta_F X_{F,ij} + \beta_G X_{G,j} 
+ \beta_{VF}(X_{V,ij}\times X_{F,ij})
+ \beta_{VG}(X_{V,ij}\times X_{G,j}) + u_{0j} + \epsilon_{ij},
\]
with $\epsilon_{ij}\sim \mathcal{N}(0,\sigma^2)$. 

For interpretability, exponentiated coefficients are reported as odds ratios (accuracy models) or time ratios (response time models).

\subsubsection*{Multiple Comparisons}
All hypothesis tests were two-sided. Holm’s method was applied to control the family-wise error rate across multiple comparisons \parencite{Holm1979}. Results are reported as effect sizes (odds ratios, time ratios, mean differences) with 95\% confidence intervals. For order effects models, we specifically tested whether version effects differed between order groups using interaction contrasts.

\section{Results}

A total of 54 participants completed the study (average age 25 years, range 18–48). Most are pursuing a bachelor’s or master’s degree, and chart literacy varied from novice to expert, with the majority reporting intermediate to advanced experience. Results are based on mixed-effects regression models that account for the repeated-measures design. Raw descriptive summaries mirrored the model estimates and are provided in \appref{app:descriptive}. 

Across the three visualization cases, modernization yielded distinct effects: Minard’s map showed clear accuracy gains from redesigned plots, Playfair’s dual-axis modern variant underperformed relative to both historical and alternative redesigns, and Nightingale’s diagram remained highly accurate while showing efficiency benefits in response time. These findings are consistent with theoretical guidance on graphical encoding and task alignment.

\subsection{Core performance}

\subsubsection{Accuracy}

\begin{figure}[H]
  \centering
  \includegraphics[width=\textwidth,height=6cm]{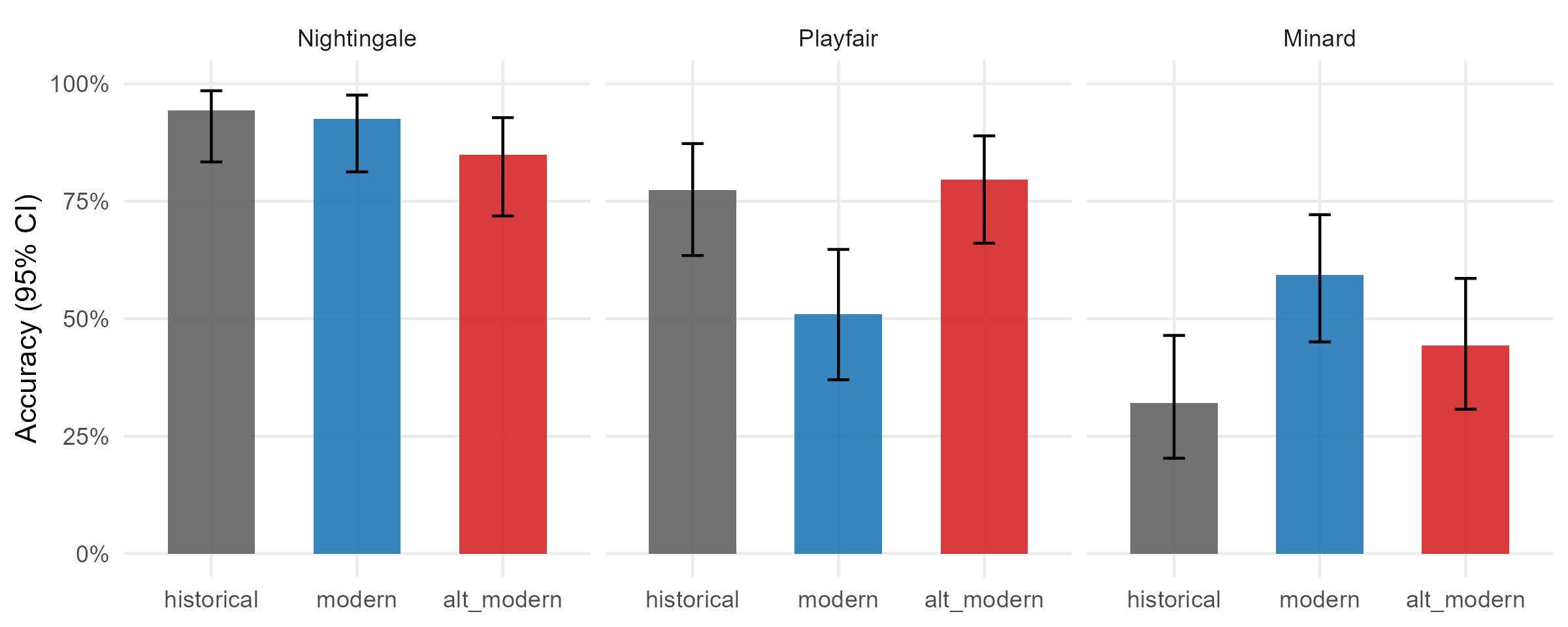}
  \caption{Model-predicted accuracy (95\% CI) by dataset and visualization version. Nightingale shows uniformly high performance across versions; Playfair’s dual-axis modern variant underperforms relative to historical and alternate modern designs; Minard shows substantial performance improvement from modern redesign.}
  \label{fig:model_acc}
  
\end{figure}

Model-based estimates from the logistic mixed-effects analysis revealed distinct patterns across datasets and visualization versions (Figure~\ref{fig:model_acc}). For Nightingale, all versions performed at near ceiling, with predicted accuracies consistently at around 90\%, indicating robustness to design changes. For Playfair, accuracy varied markedly by version: the modern dual-axis design underperformed, with predicted accuracy near 50\%, whereas both the historical and alternate modern slopegraph versions yielded substantially higher accuracy, exceeding 75\%. For Minard, the historical version showed a notably low baseline of around 30\%, but modern redesign improved performance,  reaching nearly 60\% accuracy; the alternate modern redesign showed improvements in accuracy but remained below the other modern redesign. Overall, these results suggest that while some historical charts (e.g., Nightingale) remain effective for contemporary readers, others (e.g., Playfair’s dual-axis) suffer from design limitations, and some (e.g., Minard) clearly benefit from redesign. 

\begin{figure}[H]
  \centering
  \includegraphics[width=\textwidth, height=6cm]{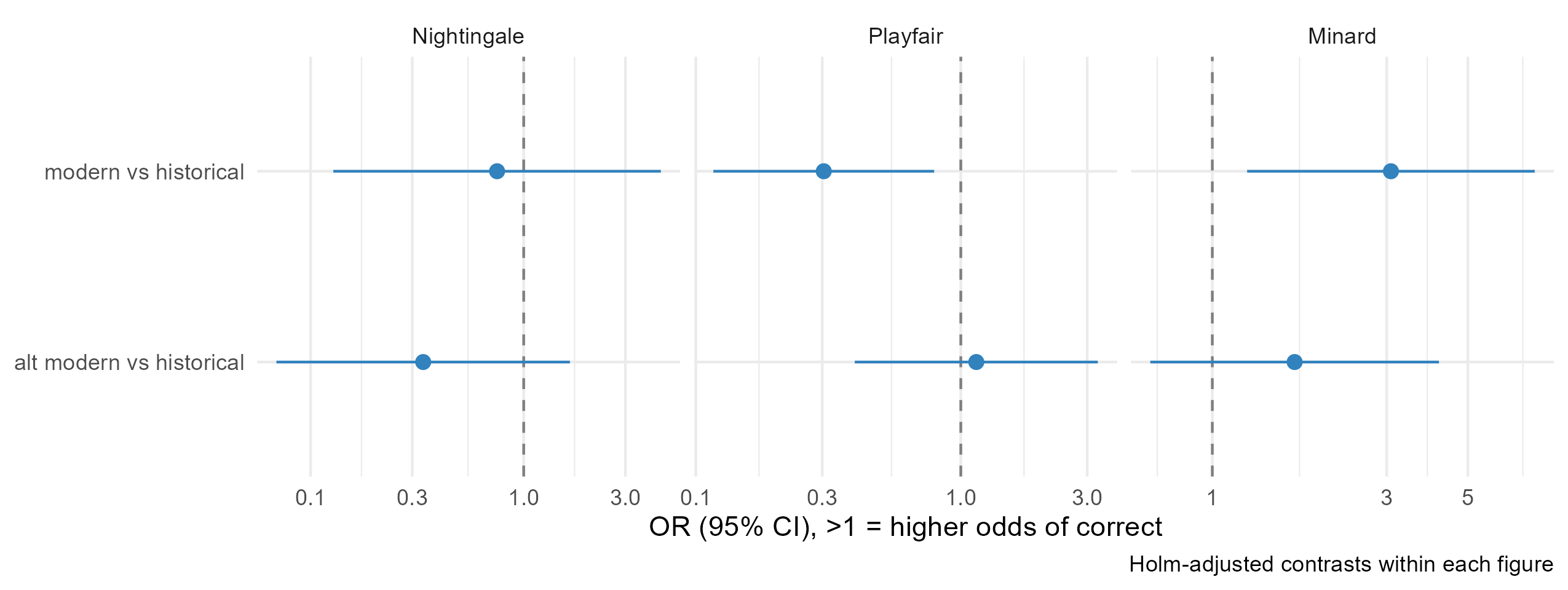}
  \caption{Odds ratios (95\% CI, log scale) comparing modern and alternate modern redesigns to historical baselines, separated by dataset. Values above 1 indicate higher odds of a correct response relative to the historical chart; values below 1 indicate lower odds.}
  \label{fig:acc2}
  
\end{figure}

To further probe differences across versions, odds ratios were estimated relative to the historical baseline for each figure (Figure~\ref{fig:acc2}). For Nightingale, the modern redesign was around the same with the historical polar diagram (OR = 1), while the alternate modern version performed worse, with an odds ratio closer to 0.3, suggesting reduced accuracy. For Playfair, the modern dual-axis chart also underperformed relative to the historical baseline (OR = 0.3), whereas the alternate modern slopegraph modestly improved performance, with odds slightly above 1. For Minard, both redesigns substantially outperformed the historical map, with the modern panel reaching odds ratios near 3 and the alternate modern design around 2, indicating clear gains in accuracy. Overall, these results show that Nightingale remains largely robust except for the weaker alternate redesign, Playfair’s accuracy is highly sensitive to design choice, and Minard benefits most strongly from redesign.

\subsubsection{Response Time}
\begin{figure}[H]
  \centering
  \includegraphics[width=\textwidth, height= 6.5cm]{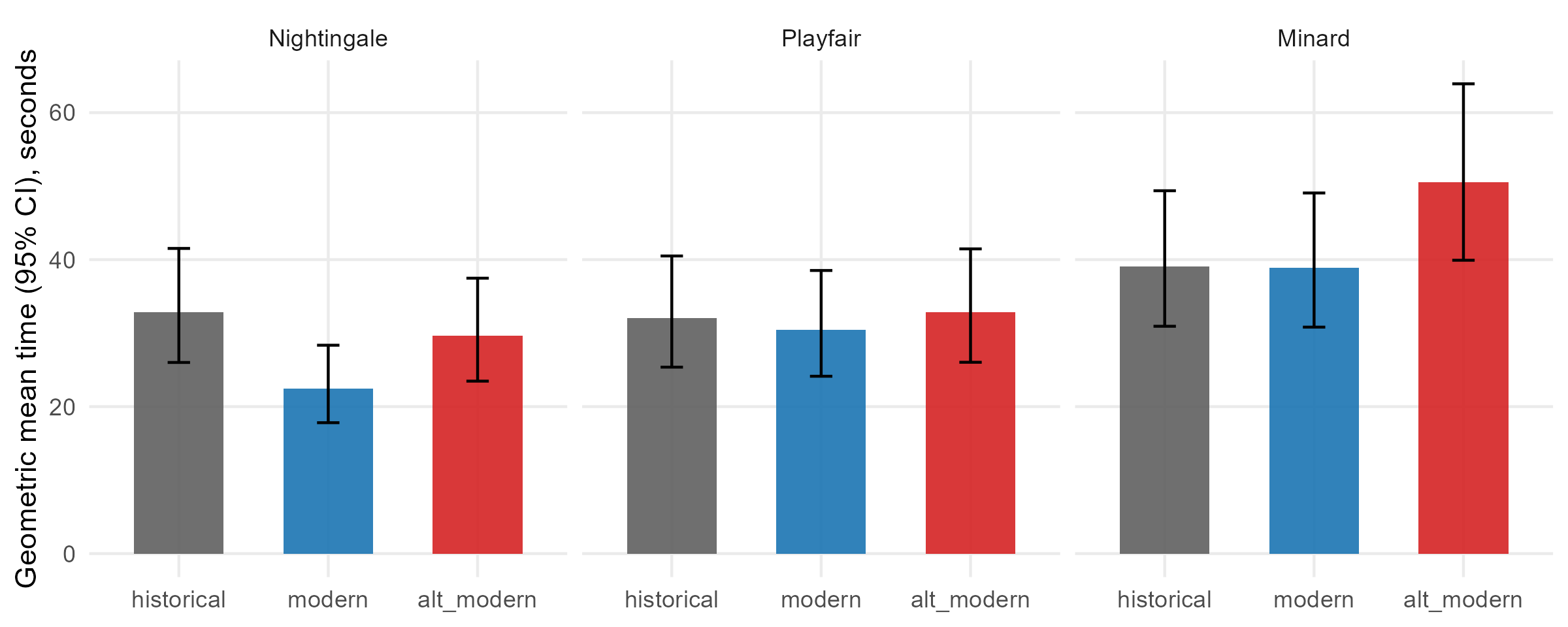}
  \caption{Model-predicted response times (geometric mean, 95\% CI) by figure and version. Nightingale redesigns produced consistently fast performance, Playfair variants showed comparable times, and Minard, especially the alternative modern redesign, resulted in the slowest responses.}
  \label{fig:model_time}
  
\end{figure}

Response times varied systematically across figures and visualization versions (Figure~\ref{fig:model_time}). For Nightingale, times were consistently low, with the modern redesign yielding slight gains in efficiency relative to both the historical polar diagram and the alternative modern heatmap. Playfair showed intermediate response times across versions, with little evidence that redesigns conferred meaningful advantages. By contrast, Minard consistently resulted in longer response times, with the alternative modern redesign producing the slowest performance of all conditions.

These results indicate that redesigns do not uniformly enhance efficiency: while the Nightingale redesign facilitated faster responses, the Minard redesign appeared to increase processing demands.

\begin{figure}[H]
  \centering
  \includegraphics[width=\textwidth, height=6.2cm]{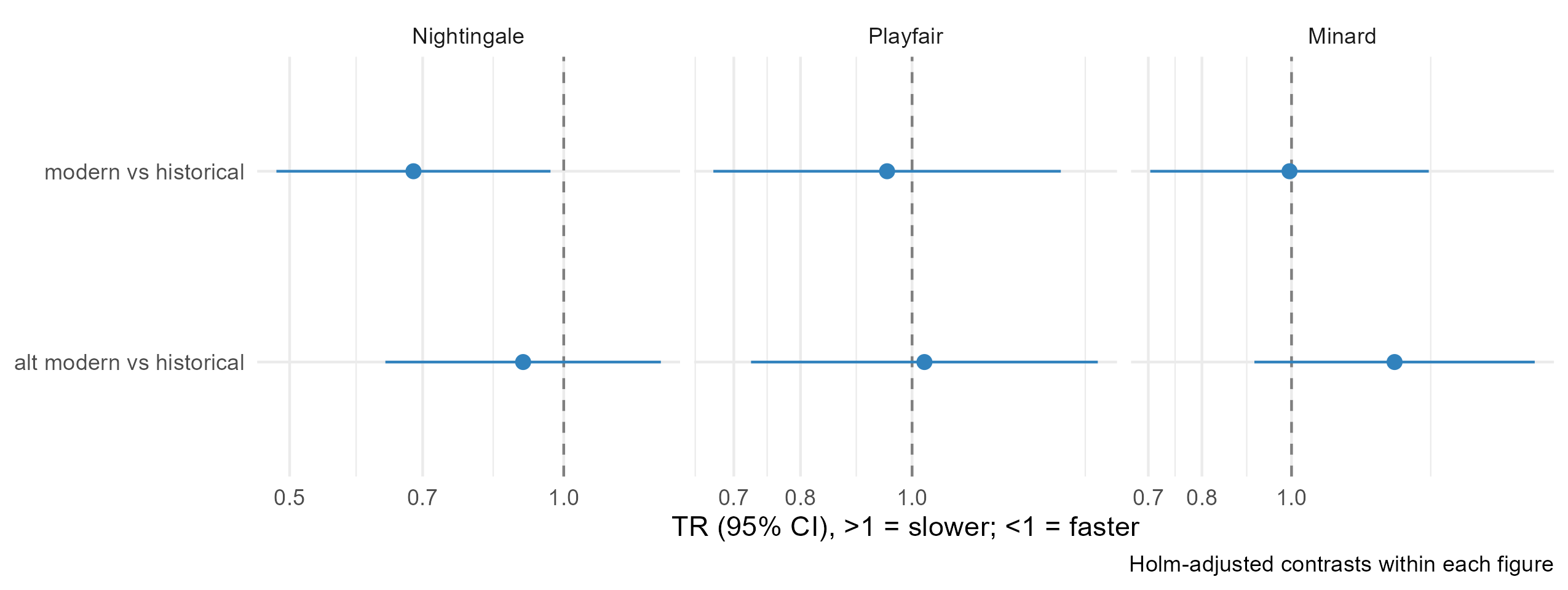}
  \caption{Time ratios (95\% CI, log scale) comparing modern and alternate modern redesigns to historical baselines, separated by dataset. Values below 1 indicate faster responses relative to the historical chart; values above 1 indicate slower responses.}
  \label{fig:time2}
  
\end{figure}

From Figure~\ref{fig:time2}, response times were evaluated as time ratios relative to the historical baseline for each figure. For Nightingale, both modern redesigns yielded faster responses than the historical polar diagram (time ratios clearly $<1$), with the modern chart showing the strongest reduction. For Playfair, neither redesign demonstrated a consistent efficiency advantage:  the modern dual-axis design produced response times essentially equivalent to the historical recreation, while the alternate modern variant suggested a slight tendency toward slower responses. For Minard, the modern redesign produced response times statistically indistinguishable from the historical version. The modern redesign and the historical recreation were essentially similar in performance (TR = 1), while the alternate modern redesign had slower response time compared with the historical plot. Taken together, these results indicate that redesigns did not systematically reduce processing time: Nightingale benefited most clearly, whereas Playfair and Minard showed little measurable efficiency gain.

\subsection{Perceived Workload (NASA-TLX)}

\begin{figure}[H]
  \centering
  \includegraphics[width=\textwidth, height=5.8cm]{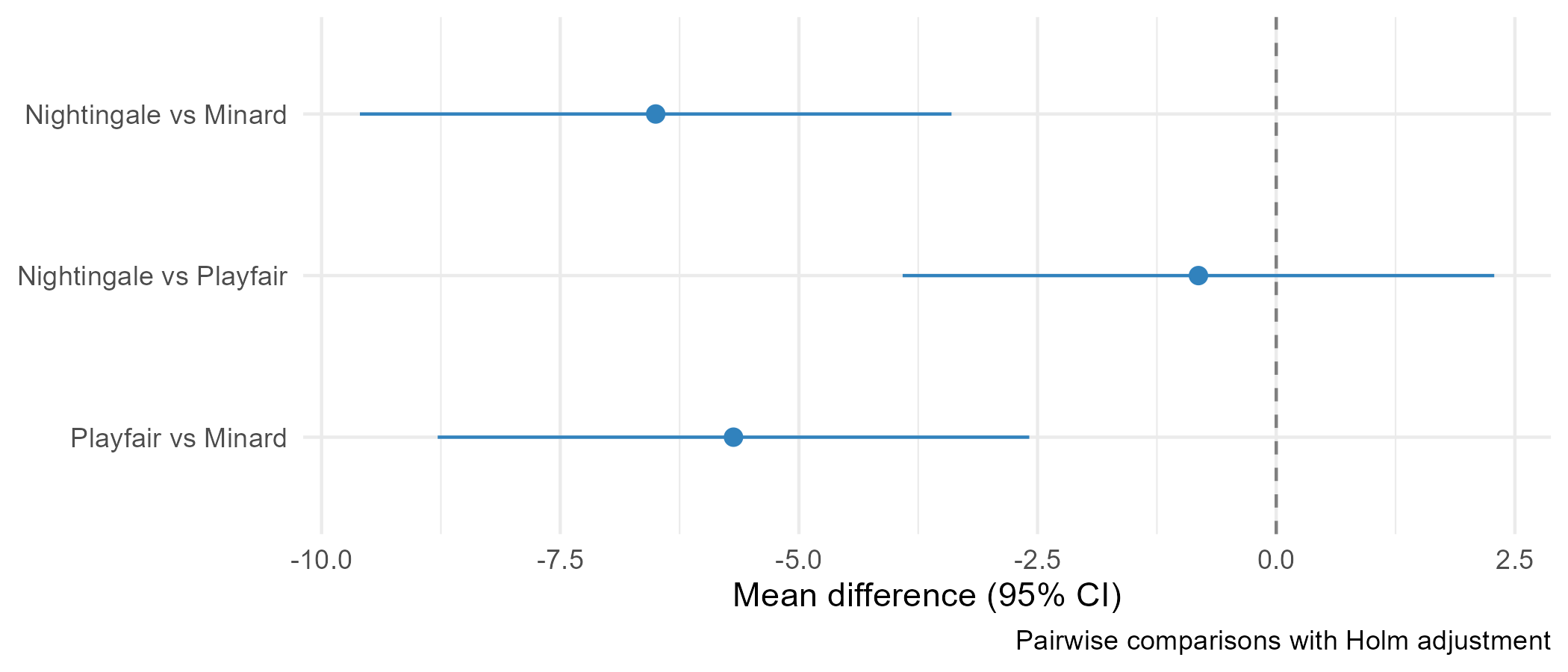}
  \caption{Pairwise differences in perceived workload (NASA–TLX, 1–10 scale) across visualizations, shown as mean differences with 95\% confidence intervals (Holm-adjusted). Negative values indicate lower workload for the first-named chart. Nightingale was consistently rated as least demanding, Playfair intermediate, and Minard most demanding.}
  \label{fig:model_workload}
  
\end{figure}

Workload comparisons indicated pronounced differences in perceived task demands across the three visualizations (Figure~\ref{fig:model_workload}). Minard was consistently rated as substantially more demanding than both Nightingale and Playfair, with mean differences of roughly six points on the NASA–TLX scale. Nightingale, by contrast, was associated with the lowest workload, aligning with its uniformly high accuracy and shorter response times. Playfair fell between, with workload ratings that were slightly higher than Nightingale but markedly lower than Minard. Ratings were collected using the NASA–TLX instrument (1–10 scale, higher values indicating greater perceived workload). These results suggest that while nineteenth-century charts can still support accurate judgments, they vary sharply in the cognitive effort they impose, with Minard in particular showing increased workload despite accuracy increases under redesign.

\subsection{Order Effects}

\begin{figure}[H]
  \centering
  \includegraphics[width=\textwidth, height = 5.5cm]{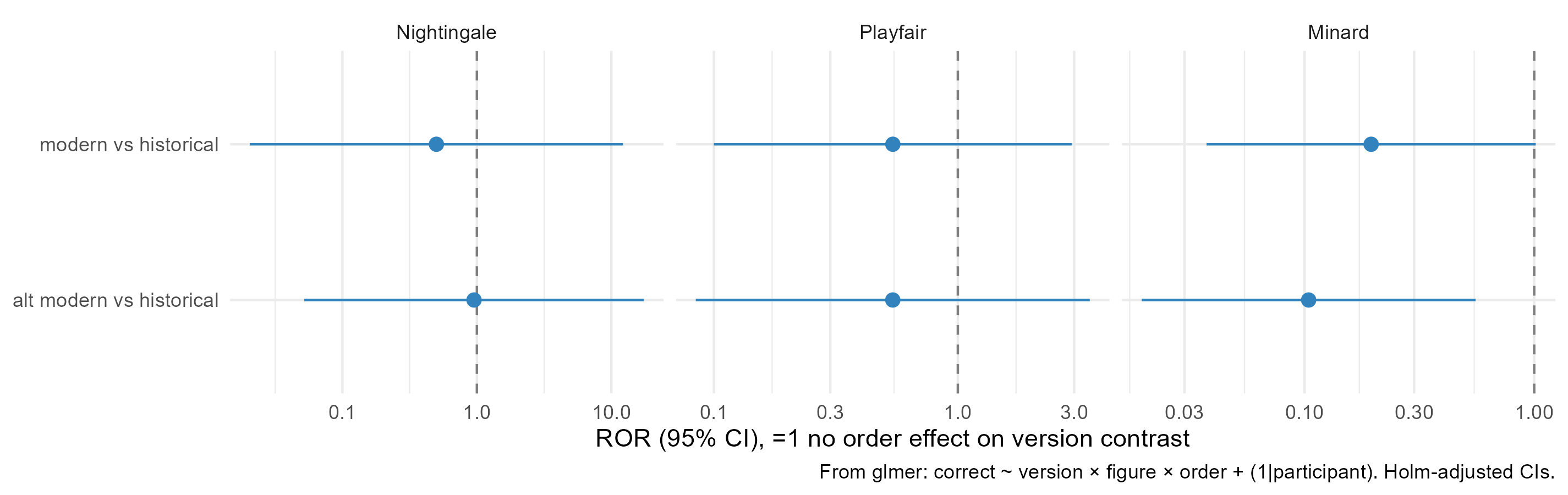}
  \caption{Odds ratios (95\% CI, log scale) for accuracy contrasts of modern and alternative modern designs relative to the historical baseline, estimated separately for Nightingale, Playfair, and Minard. Values above 1 indicate higher odds of a correct response. Results show stable version effects across groups, with Nightingale robust to redesign, Playfair disadvantaged under both redesigns, and Minard remaining difficult despite alternative versions.}
  \label{fig:order1}
  
\end{figure}

Analysis of odds-ratio contrasts relative to the historical baseline showed that order did not alter version effects (Figure~\ref{fig:order1}). For Nightingale, performance remained uniformly high, with the modern redesign slightly below the historical polar diagram (OR = 0.7) and the alternative modern design performing at parity (OR = 1). For Playfair, both the modern redesign and the alternative modern redesign underperformed relative to the historical baseline (OR = 0.65 in both cases), indicating a consistent disadvantage regardless of order. For Minard, odds ratios appeared below 1 in the order-stratified model because accuracy was lower when Minard was presented first; however, the version-by-order interaction was not significant, indicating that order did not materially moderate the redesign advantage observed in the primary model. Taken together, these findings indicate that the observed version effects, Nightingale’s consistent effectiveness, Playfair’s persistent weakness under redesign, and Minard’s redesign-related improvements, reflect design properties rather than order effects.

\begin{figure}[H]
  \centering
  \includegraphics[width=\textwidth]{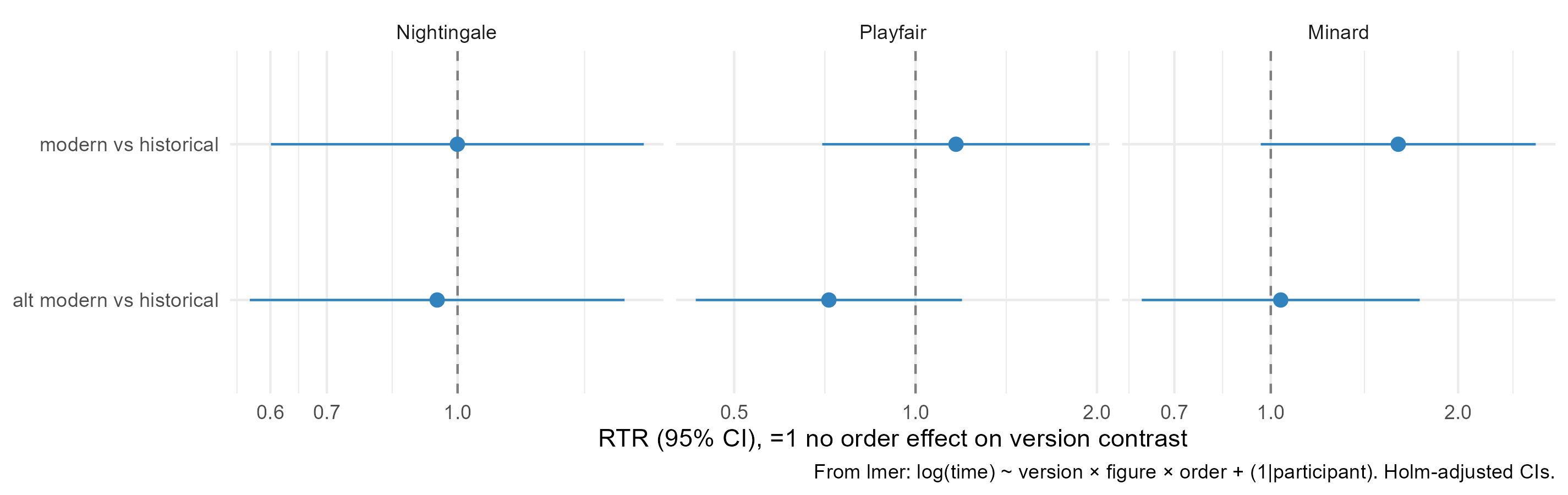}
  \caption{Time ratios (95\% CI, log scale) for response-time contrasts of modern and alternative modern designs relative to the historical baseline, estimated separately for Nightingale, Playfair, and Minard. Ratios below 1 indicate faster responses than historical. Results show stable patterns across orders, with Nightingale largely unaffected, Playfair’s slopegraph faster but dual-axis slower, and Minard remaining slower regardless of redesign.}
  \label{fig:order2}
  
\end{figure}

Time-ratio contrasts also revealed little evidence of order effects (Figure~\ref{fig:order2}). For Nightingale, response times were essentially unchanged across designs, with the modern version at near equivalence and the alternative modern redesign showing only a marginal speedup relative to the historical polar diagram. For Playfair, clear design differences emerged: the modern redesign was slower than the historical baseline, whereas the alternative modern redesign was faster at 0.7. For Minard, response times were increased for both redesigns, with the modern redesign particularly slow and the alternative redesign only modestly faster, though still not below the historical map. Overall, these effects were consistent across orders, suggesting that the timing differences reflect design properties rather than sequencing or fatigue.

\section{Discussion}

\subsection{Summary of Findings}

The central aim of this study was to assess whether nineteenth-century graphics remain effective for contemporary readers. In some cases, the original designs are effective: Nightingale’s polar diagram achieved near-ceiling accuracy and low workload regardless of redesign, indicating that her encoding choices remain well suited to contemporary interpretation. In other cases, however, historical forms reveal limitations. Playfair’s modern plot consistently underperformed relative to both the historical recreation and the modern alternative, showing that certain encoding strategies that may have been useful in the nineteenth century are now less effective given today’s expectations and cognitive demands. Minard’s map highlights a third pattern: while the historical form yielded very low accuracy and high workload, modern redesigns substantially improved performance without introducing additional time costs.

These findings indicate that nineteenth-century charts do not uniformly “work” in modern contexts. Some (e.g., Nightingale) remain both accurate and efficient, while others (e.g., Playfair, Minard) require redesign to meet contemporary standards of usability. More broadly, the study shows that the persistence of historical charts depends less on their age than on the perceptual adequacy of their designs: clear encodings remain effective, while cognitively demanding or complex ones can benefit from adaptation.

\subsection{Implications for Visualization Design}

As outlined in Methods, the redesigns contrasted historical encodings with alternatives based on perceptual principles. Here we discuss how those design choices influenced performance outcomes and what broader lessons they suggest for visualization design.

The Minard case illustrates how design choices shape the legibility of complex narratives \parencite{robinson1967minard}. 
The flow-map recreation was evocative but relied on width along curved paths, a weak perceptual 
channel that contributed to low accuracy. Both the small-multiples lines and the binned bars 
replaced width with position or length on a common scale, producing increased accuracy. 
Yet these redesigns also raised workload, as readers had to scan axes and panels rather than 
follow a single integrated story. This suggests that redesigns can increase precision but at 
the cost of interpretive ease, and that scaffolds such as reference labels or aligned axes are 
needed to mitigate added effort.

Nightingale’s diagrams, by contrast, show how strong task alignment and redundancy can sustain 
usability even when weaker encodings are involved. The polar-area diagram used radial area, 
but its cyclic structure, simple palette, and emphasis on seasonal patterning supported 
near-ceiling accuracy and low workload. The stacked bars and heatmap offered incremental 
gains, yet the historical form already matched the analytic questions. This pattern highlights 
that when redundancy aligns with analytic goals, perceptually suboptimal channels can still be 
effective, and modernization yields only modest benefits.

The Playfair case demonstrates the risks of modernizations that complicate rather than clarify. The historical two-line chart allowed accurate comparisons on a shared baseline, and the stacked area chart variant emphasized overall trade volume and composition while maintaining interpretability. In contrast, the dual-axis redesign introduced competing scales, reducing accuracy without efficiency gains. This shows that not all modernizations add value, and that adding scales without separability can undermine performance.

Across these cases, broader principles emerge:
\begin{itemize}
    \item Use aligned positional designs for precision. Accuracy was highest when values shared a common scale.
    \item Reduce channel interference through separation. Facets and panels proved more effective than overloaded composite displays.
    \item Leverage redundancy when it supports the task. Nightingale’s success shows how aligned redundancy can aid interpretation.
    \item Balance accuracy with efficiency. Redesigns like Minard’s lines improved precision but required extra effort that could be eased through light scaffolding.
\end{itemize}

For designers adapting historical graphics today, these results suggest a pragmatic approach: 
preserve original forms when they already align with task demands, avoid redesigns that add 
complexity without clarity, and when modification is warranted, favor positional encodings and 
separable panels while adding minimal scaffolds to maintain efficiency.

These findings can be situated within established frameworks for visualization design. In Munzner’s nested model \parencite{Munzner2014VAD}, Nightingale’s polar diagram shows strong alignment at the problem and abstraction levels: seasonal mortality patterns matched the cyclic form, so even a weaker encoding channel (radial area) remained effective. Playfair’s dual-axis redesign illustrates a mismatch between abstraction (a single comparison of imports and exports) and encoding (separate axes), leading to reduced usability. Minard’s flow map highlights trade-offs within the encoding level: replacing width with position improved accuracy but increased workload, suggesting the need for scaffolds such as labels or aligned axes to maintain efficiency.

Viewed through Lam et al.’s evaluation taxonomy \parencite{Lam2012SevenScenarios}, this study contributes to the “effectiveness” branch, providing controlled evidence of how classic forms perform under modern tasks. Together, these results show that historical graphics exemplify enduring principles of visualization design: when task alignment and redundancy are present, original forms remain effective; when abstraction and encoding conflict, redesign is warranted.

\subsection{Limitations and Future Directions}

Several limitations should be acknowledged. First, the participant sample was relatively small and skewed toward younger, highly educated individuals with some prior chart experience, which may limit generalizability to broader populations. Second, demographic imbalance was present between Group A and Group B, particularly in gender distribution. Gender was not included as a factor in the statistical models, since the study was not designed to test demographic effects; nevertheless, such imbalances may have introduced additional variability into the results. Third, the study focused on only three canonical nineteenth-century graphics, and the redesigns represent specific design choices among many possible alternatives. Future work could expand to a broader range of historical designs and systematically vary redesign strategies. 

In addition, we included only four NASA–TLX subscales, as the tasks imposed minimal physical or performance demands. While appropriate for the current setting, it may not capture additional aspects of workload that could be relevant in other contexts. Finally, the question-answering tasks focused on lookup and inference, which provide clear benchmarks for accuracy and efficiency but may not capture more open-ended, exploratory, or narrative uses of visualization.

Future research could incorporate a broader set of historical charts from diverse domains to test whether the observed patterns generalize. In addition, redesign strategies could be systematically manipulated—for example, by varying color, interaction, or annotation—to identify which features most enhance usability. Finally, expanding beyond static question-answering tasks to include exploratory or narrative use cases would provide a richer understanding of how historical and modern designs support real-world reasoning.

\section*{Acknowledgments}

The author thanks the participants for their time and effort in completing the study. The author also acknowledges the use of OpenAI's ChatGPT to assist with language polishing and LaTeX formatting. All study design, data analysis, and interpretations were conducted independently by the author.

\newpage
\appendix
\renewcommand{\thesection}{Appendix \Alph{section}}
\renewcommand{\thesubsection}{\thesection.\arabic{subsection}}

\section{Study Materials}

\subsection{Visualization Designs}\label{app:vis}

\begin{figure}[ht]
  \centering
  \includegraphics[width=0.45\textwidth]{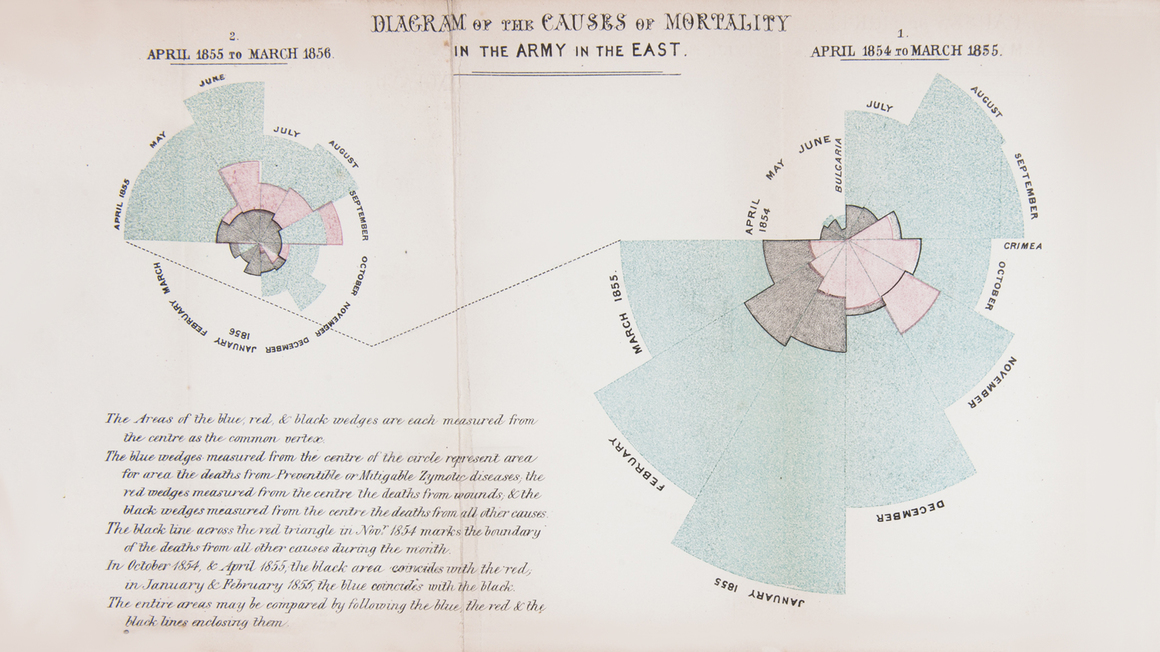}
  \includegraphics[width=0.45\textwidth, height = 7cm]{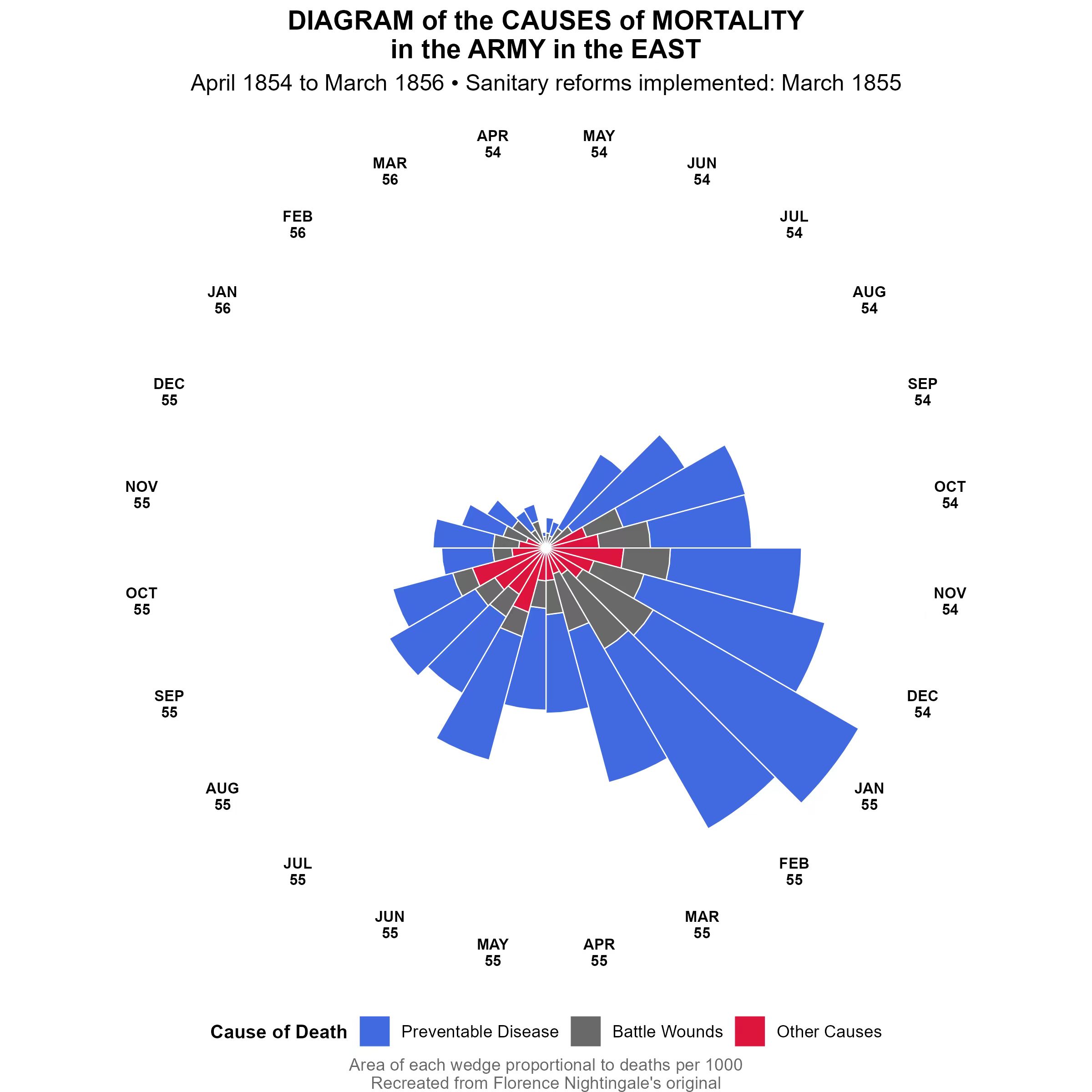}
  \caption{Nightingale original chart (left) \parencite{Nightingale1858} and historical recreation used in the study (right).}
\end{figure}

\begin{figure}[ht]
  \centering
  \includegraphics[width=0.45\textwidth]{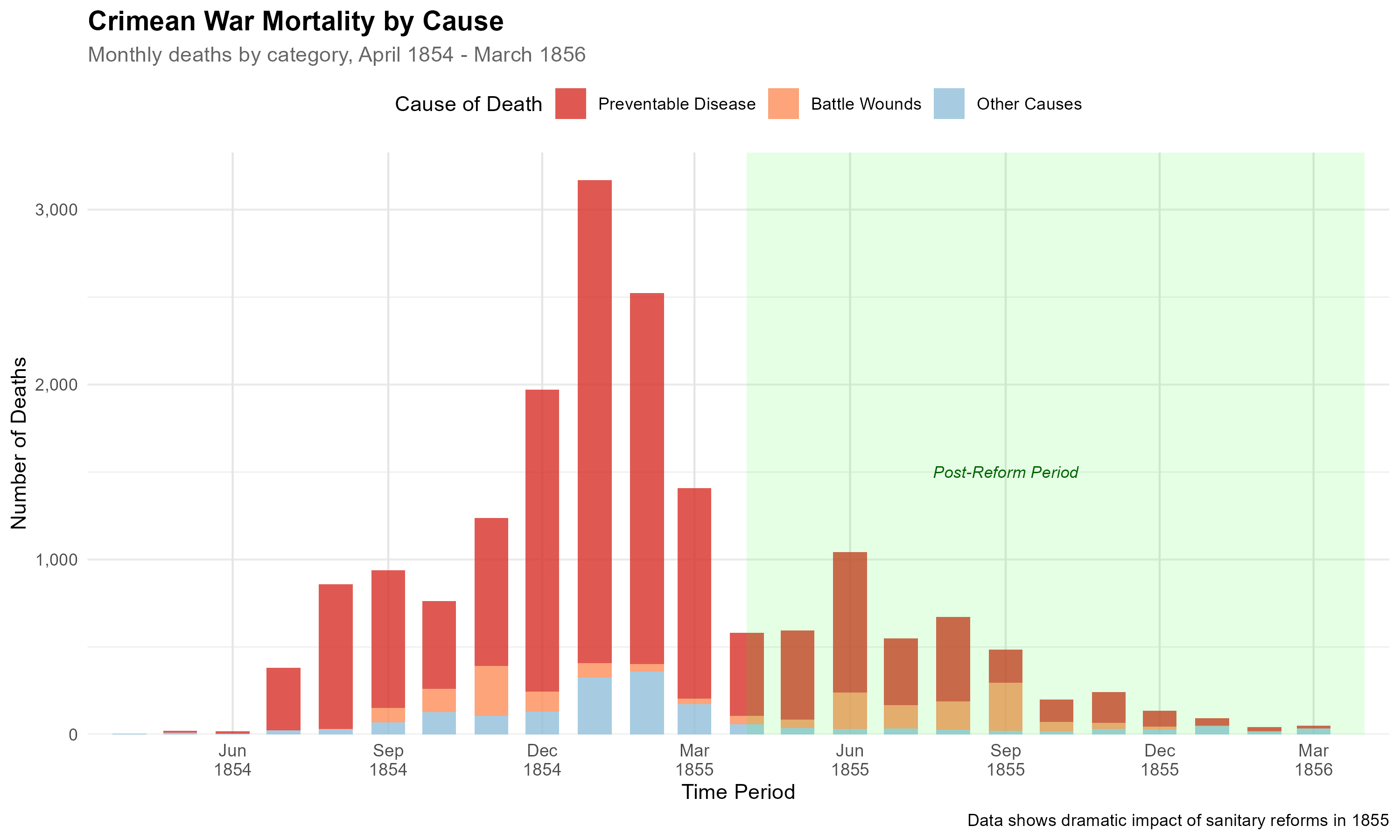}
  \includegraphics[width=0.45\textwidth]{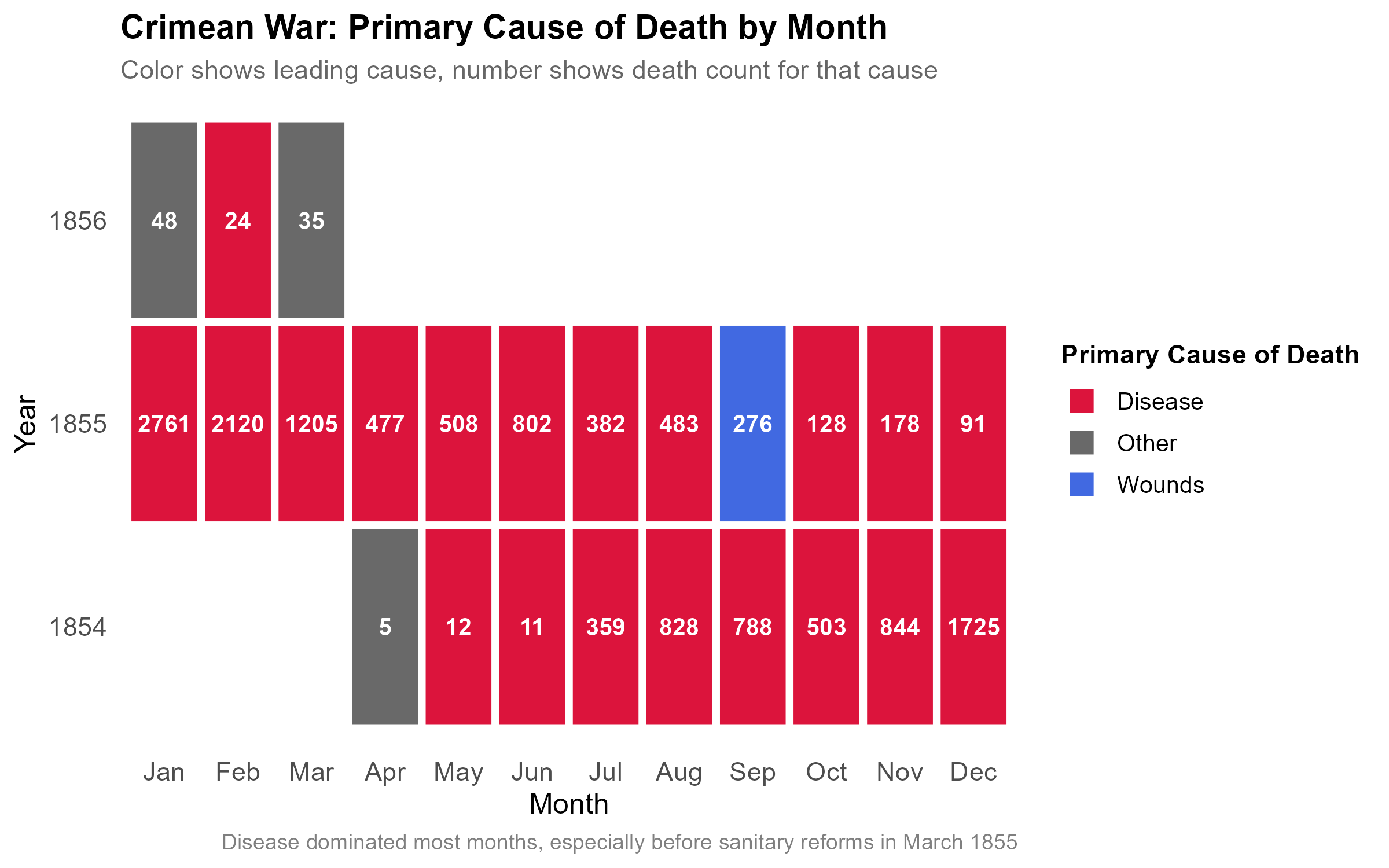}
  \caption{Nightingale modern design(left) and alternative modern design(right).}
\end{figure}

\begin{figure}[ht]
  \centering
  \includegraphics[width=0.45\textwidth, height = 4cm]{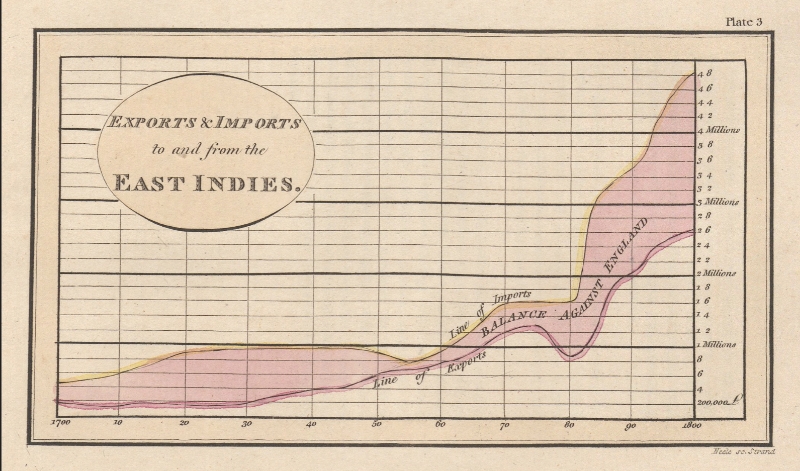}
  \includegraphics[width=0.45\textwidth, height = 4cm]{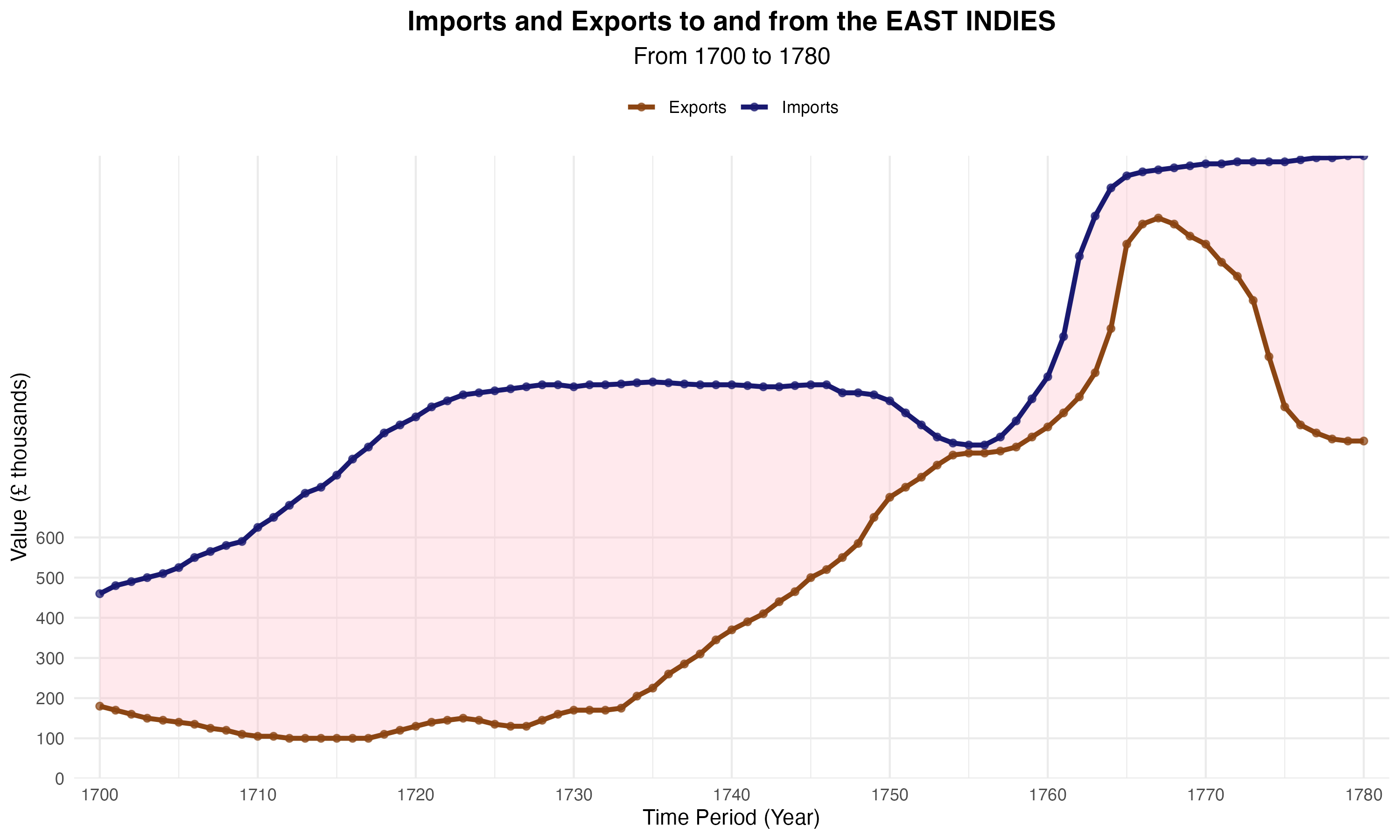}
  \caption{Playfair original chart (left) \parencite{Playfair1786} and historical recreation used in the study (right).}
\end{figure}

\begin{figure}[ht]
  \centering
  \includegraphics[width=0.45\textwidth, height= 6cm]{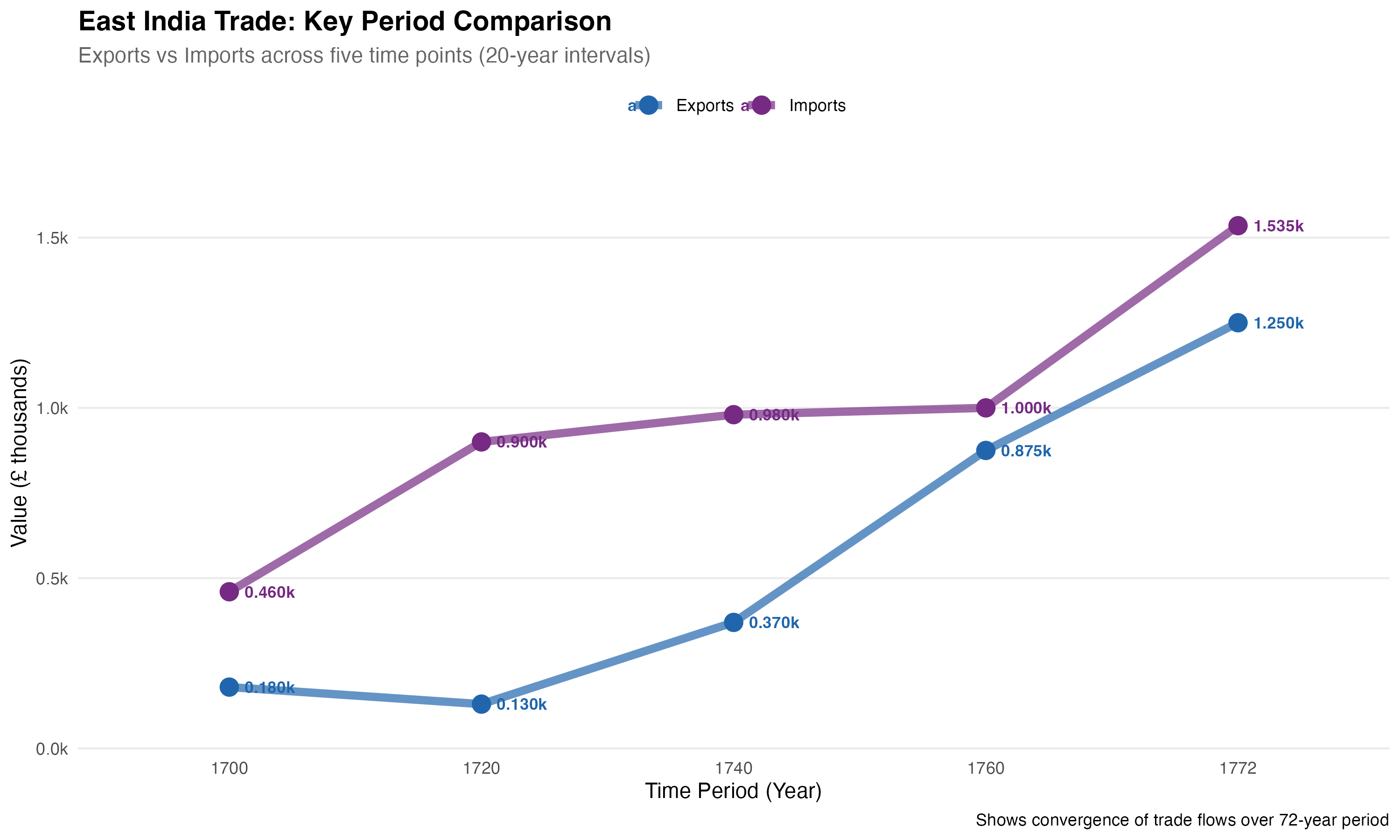}
  \includegraphics[width=0.45\textwidth]{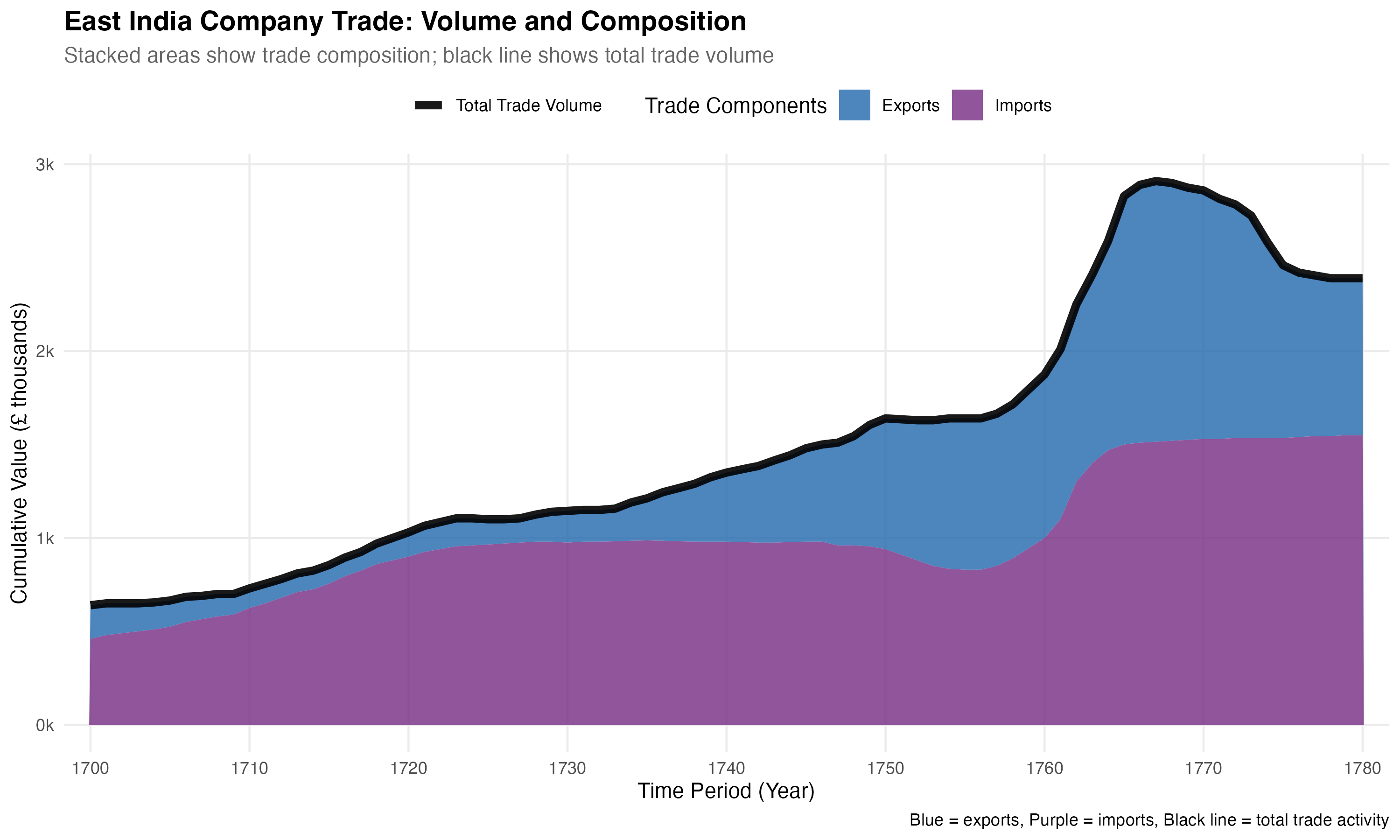}
  \caption{Playfair modern design (left) and alternative modern design (right).}
\end{figure}

\begin{figure}[ht]
  \centering
  \includegraphics[width=0.45\textwidth, height = 5cm]{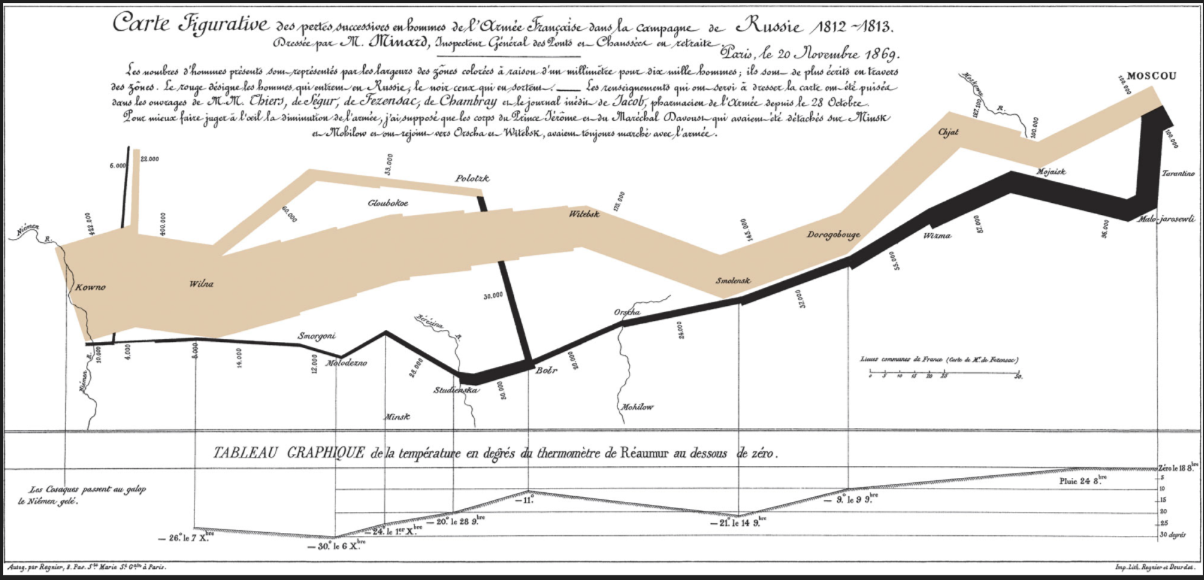}
  \includegraphics[width=0.45\textwidth]{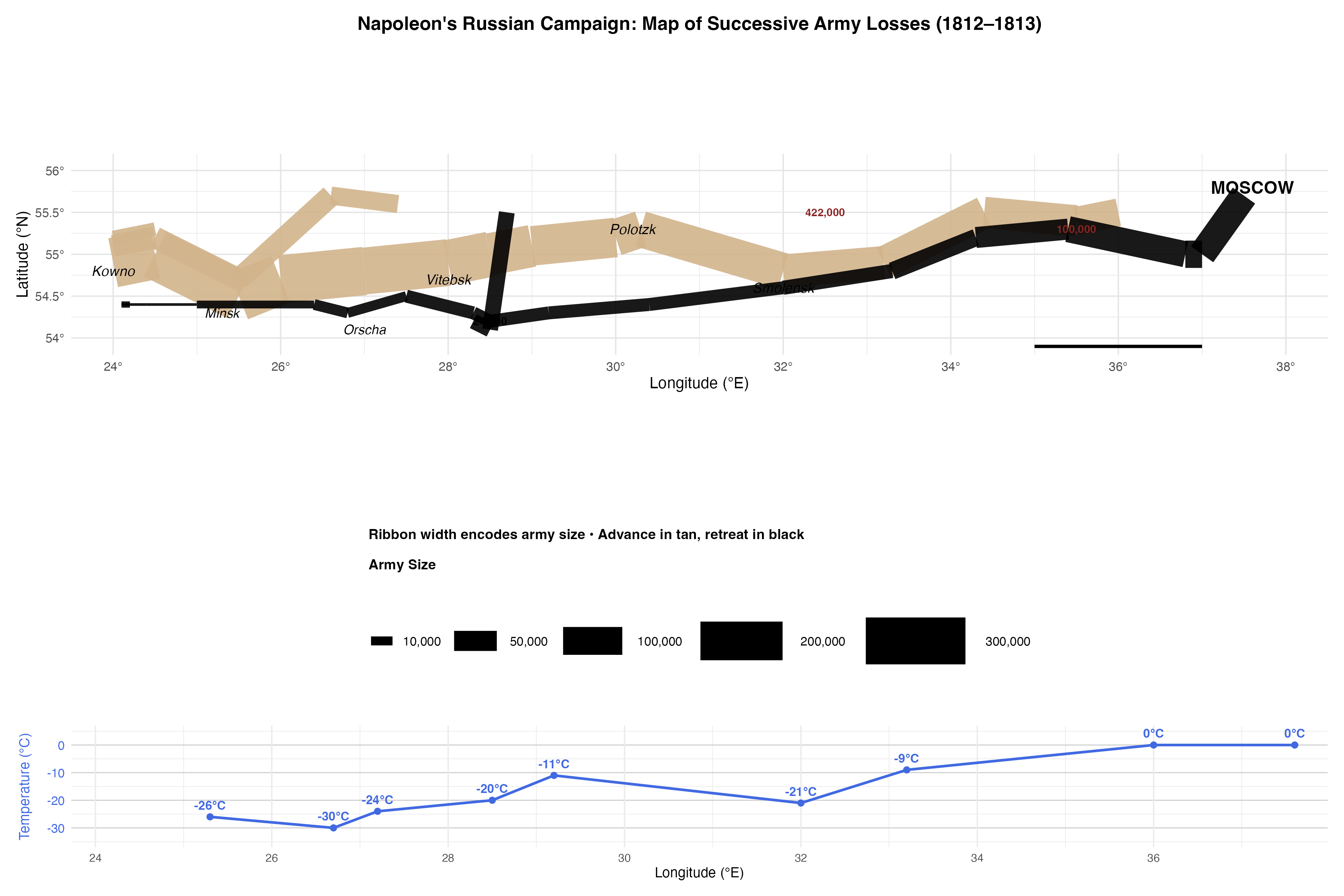}
  \caption{Minard original chart (left) \parencite{Minard1869} and historical recreation used in the study (right).}
\end{figure}

\begin{figure}[ht]
  \centering
  \includegraphics[width=0.45\textwidth, height = 5cm]{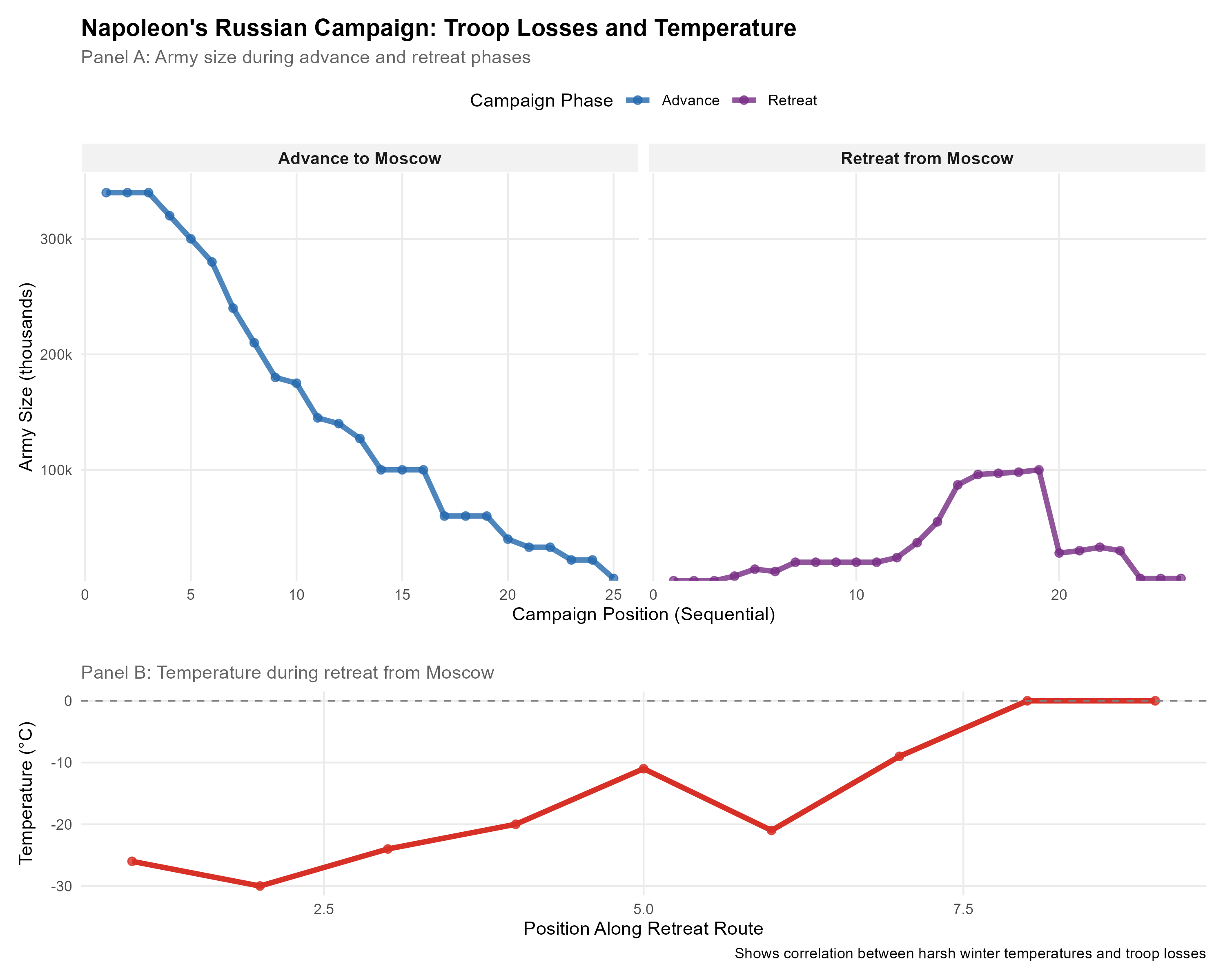}
  \includegraphics[width=0.45\textwidth]{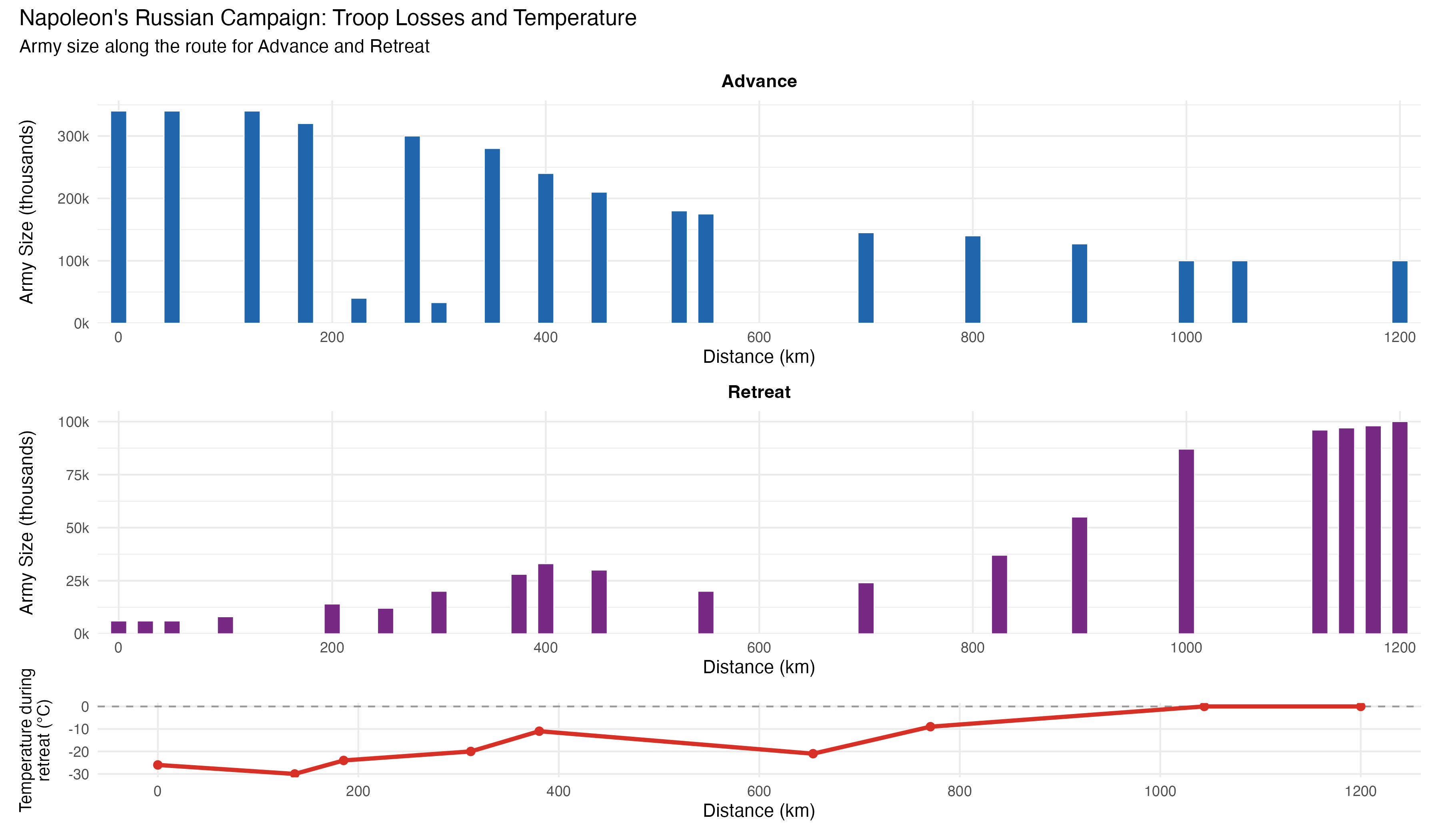}
  \caption{Minard modern design(left) and alternative modern design(right).}
\end{figure}

\clearpage

\subsection{Survey and Administration Interface}\label{app:survey}

\begin{figure}[H]
  \centering
  \includegraphics[width=\textwidth]{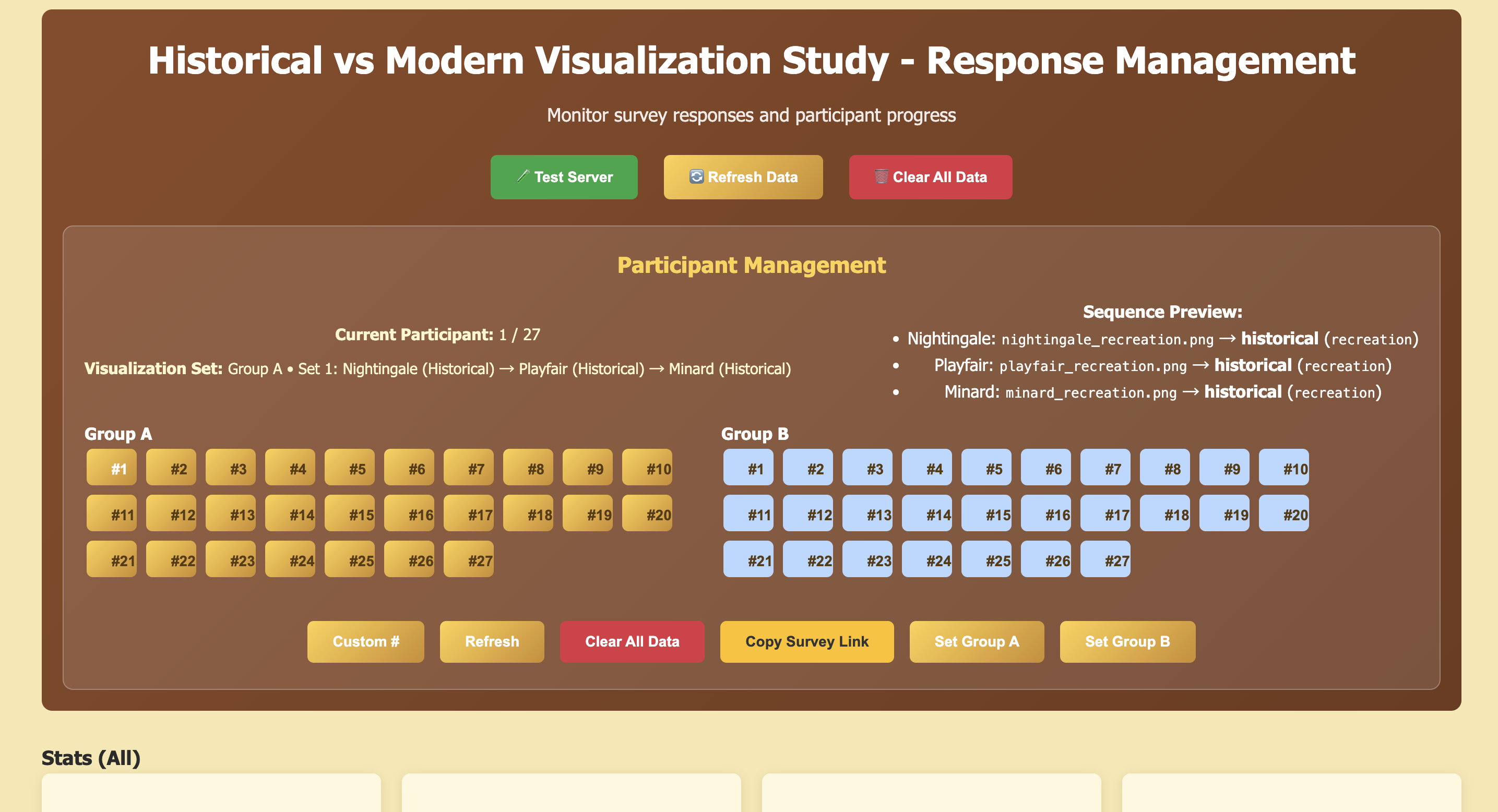}
  \caption{Administrator dashboard for session management, allowing experimenters to assign participants to Group A or B, control sequence order, and track overall study progress.}
  \label{fig:acc}
\end{figure}

\begin{figure}[H]
  \centering
  \includegraphics[width=\textwidth, height = 8cm]{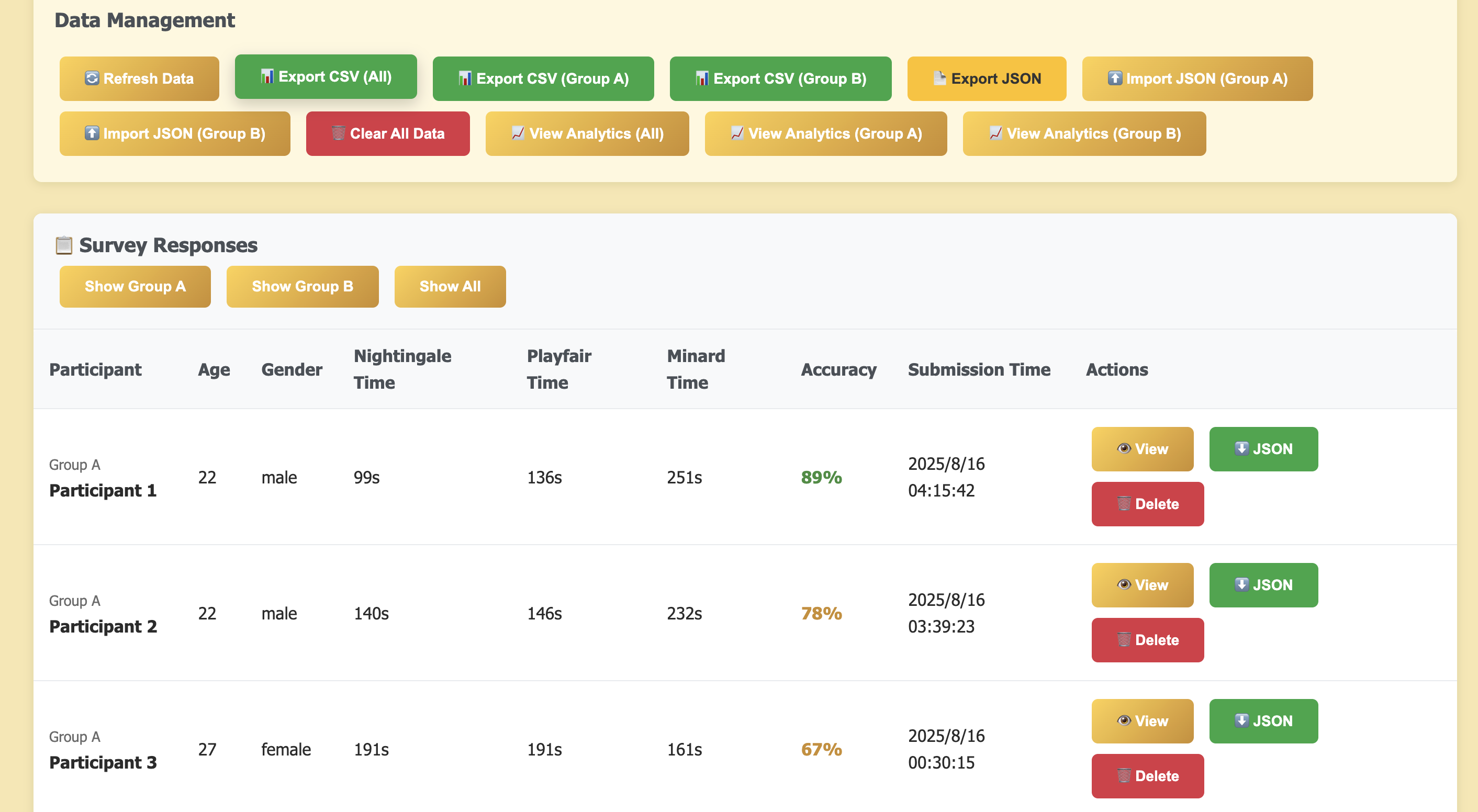}
  \caption{Administrator dashboard for performance monitoring, showing detailed participant responses including task times, response accuracy and NASA-TLX scores.}
  \label{fig:acc}
\end{figure}

\begin{figure}[H]
  \centering
  \includegraphics[width=0.45\textwidth, height = 15cm]{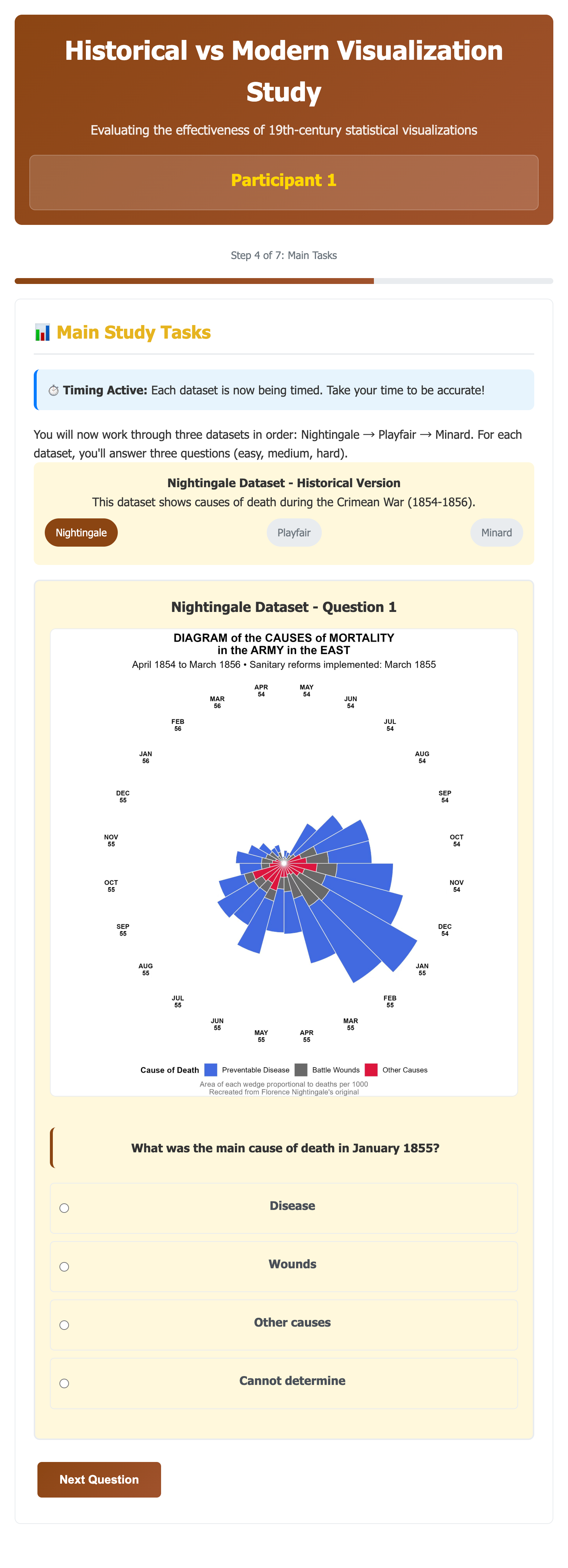}
  \includegraphics[width=0.45\textwidth]{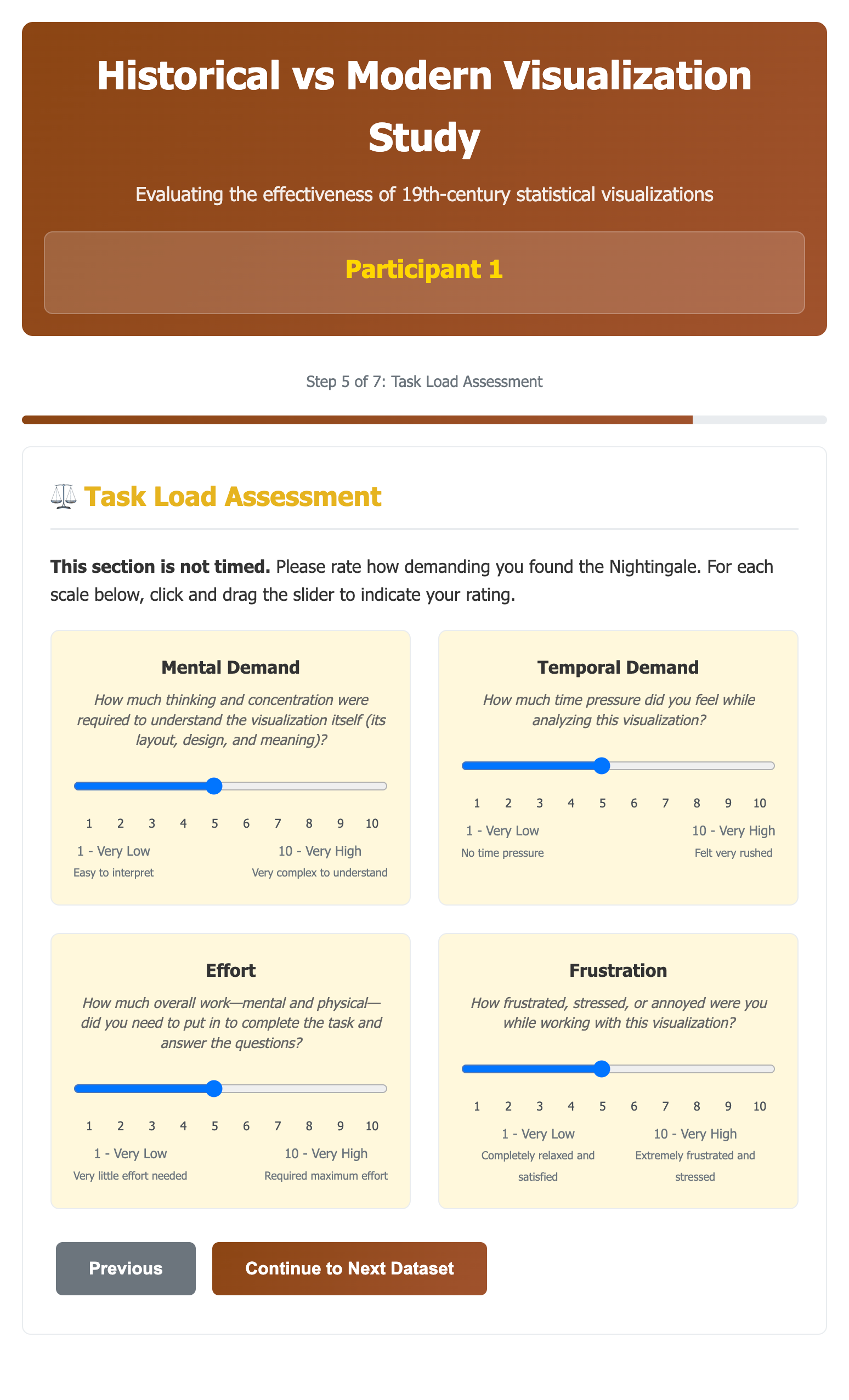}
  \caption{Example survey interface for participants. The left panel shows a main study task, in which participants answered multiple-choice questions based on historical and redesigned visualizations (here, Nightingale’s recreation). The right panel illustrates the NASA-TLX workload assessment, where participants rated mental demand, temporal demand, effort, and frustration after each dataset.}
\end{figure}

\section{Question Formatting Protocol}\label{app:questions}

To ensure comparability across datasets and visualization versions, all task questions were developed following a standardized question-formatting protocol. This protocol defined difficulty levels, answer structure, and wording style:

\begin{itemize}
    \item \textbf{Difficulty Levels:}
    \begin{itemize}
        \item \textbf{Easy} – A straightforward lookup from a single data point (basic data reading).
        \item \textbf{Medium} – A targeted pattern search, requiring recognition or comparison across values.
        \item \textbf{Hard} – An interpretation or inference task, drawing on cause–effect relationships or broader trends in the chart.
    \end{itemize}

    \item \textbf{Answer Structure:}  
    Each question provided exactly four multiple-choice options.  
    Where appropriate, one option was \emph{``Cannot determine''} to account for cases where the visualization did not explicitly support a clear answer.

    \item \textbf{Wording and Parallel Phrasing:}  
    To minimize cognitive load and improve comparability, questions across datasets were phrased in a parallel structure:
    \begin{itemize}
        \item Q1 begins with \emph{``What was…''} (single-point lookup).  
        \item Q2 begins with \emph{``In which…''} or a similar targeted prompt (pattern search).  
        \item Q3 begins with \emph{``What does…''} or equivalent inference phrasing (interpretation).  
    \end{itemize}
\end{itemize}

This protocol ensured that differences in participant performance could be attributed to the visualization design and dataset rather than to wording variability in task questions.

\section{Raw Descriptive Plots}\label{app:descriptive}

\subsection{Accuracy and Response Time}
Descriptive results in Figure~\ref{fig:acc} show that Nightingale is uniformly high across versions, with the historical polar and modern bar charts nearly indistinguishable, and only a modest reduction for the alternative modern heatmap.

\begin{figure}[H]
  \centering
  \includegraphics[width=\textwidth]{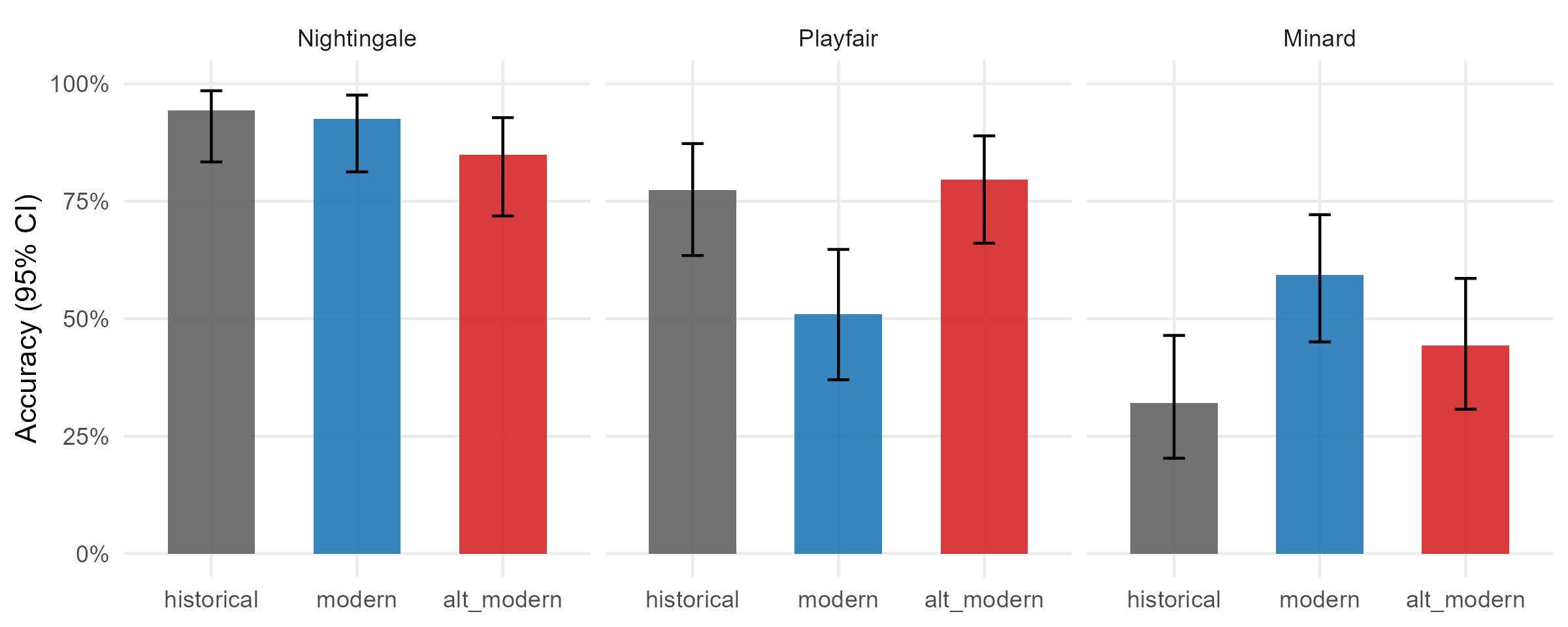}
  \caption{Descriptive Accuracy (95\% CI) by figure and version. Nightingale is uniformly high; Playfair’s dual-axis modern variant underperforms; Minard benefits most from modernization.}
  \label{fig:acc}
\end{figure}

\begin{figure}[H]
  \centering
  \includegraphics[width=\textwidth]{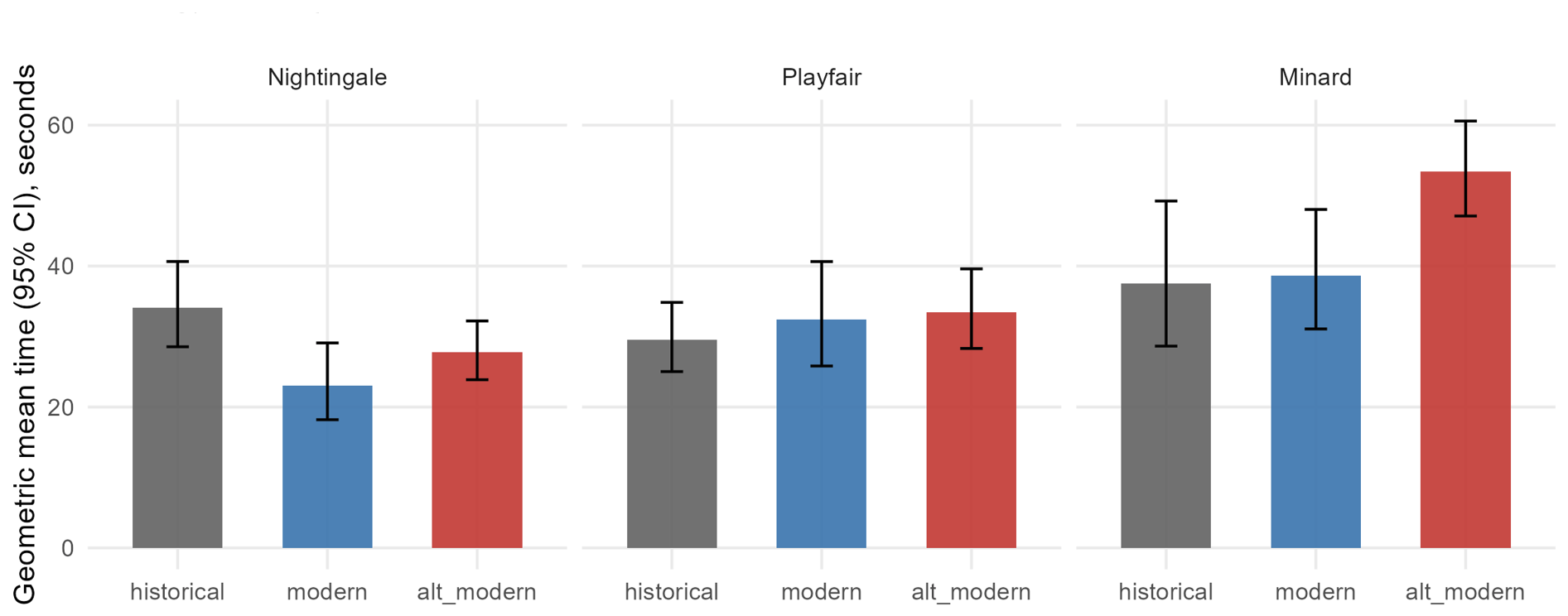}
  \caption{Response time (geometric mean, 95\% CI) by figure and version. Nightingale modern is fastest; Playfair versions are similar in time; Minard alternate modern is slowest.}
  \label{fig:time}
\end{figure}

\subsection{Perceived workload (NASA--TLX)}

\begin{figure}[H]
  \centering
  \includegraphics[width=\textwidth, height = 5cm]{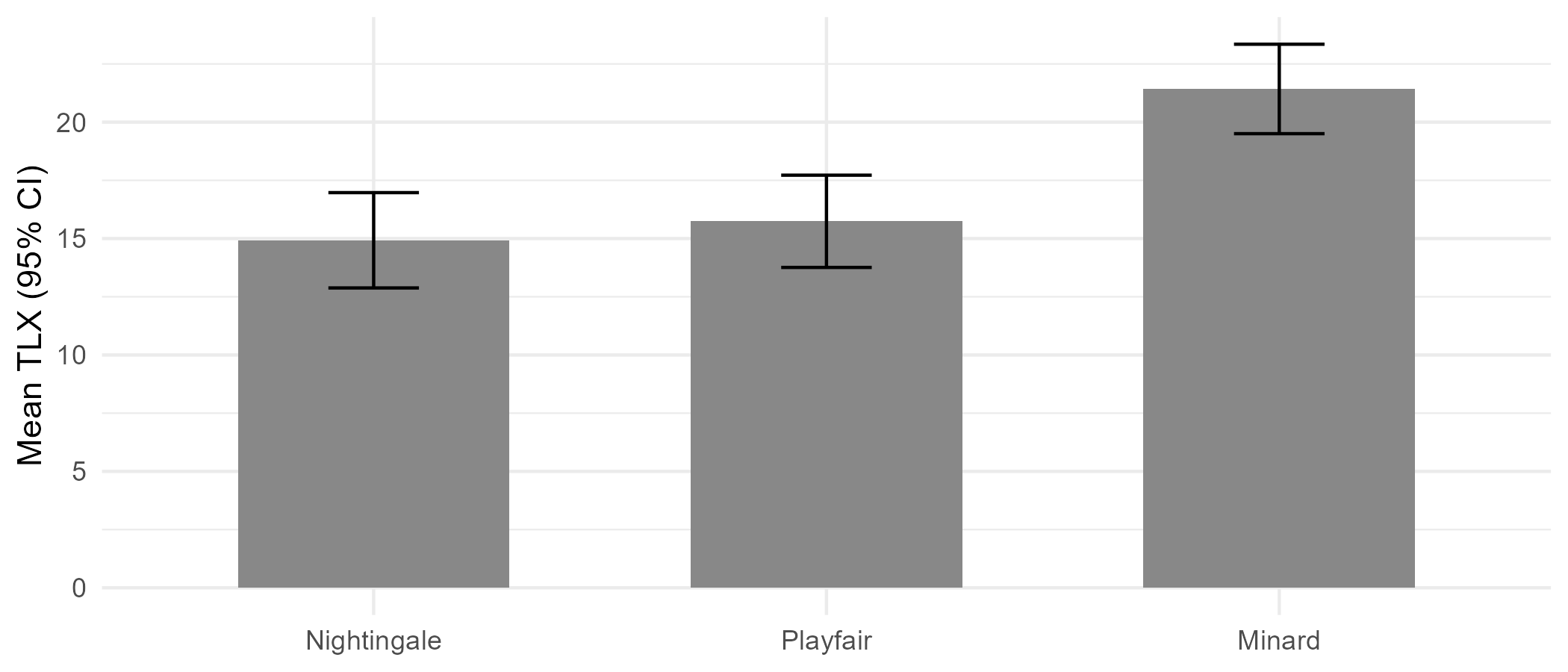}
  \caption{Perceived workload (NASA-TLX) by figure: Mean TLX by figure (95\% CI). Minard is hardest; Nightingale is easiest.}
  \label{fig:tlx-fig}
\end{figure}

\begin{figure}[H]
  \centering
  \includegraphics[width=\textwidth, height = 5cm]{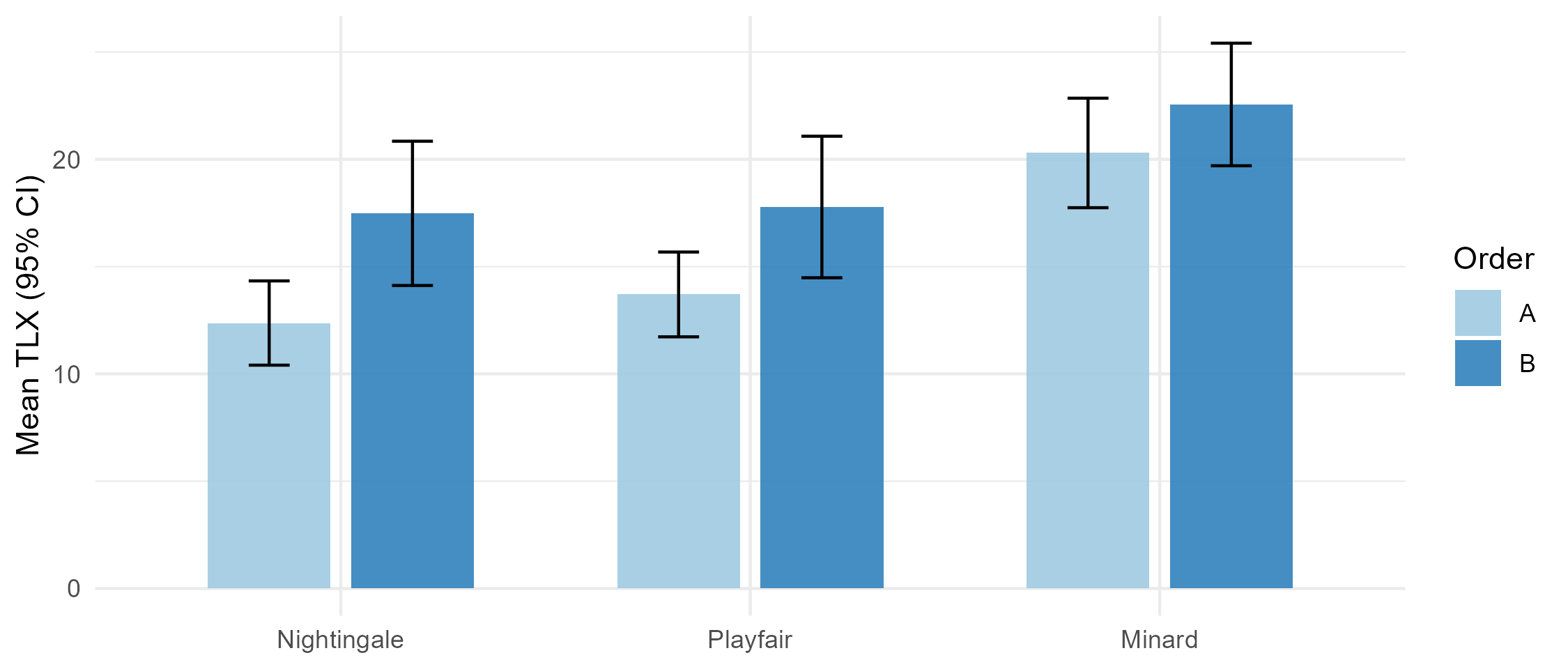}
  \caption{Perceived workload by figure and order group: Mean TLX by figure and order group (95\% CI). Starting with Minard raises perceived workload across the session.}
  \label{fig:tlx-group}
\end{figure}

\subsection{Order effects}
Group A saw Nightingale→Playfair→Minard; Group B saw Minard→Playfair→Nightingale. This separates design from session order.

\begin{figure}[H]
  \centering
  \includegraphics[width=\textwidth, height = 8cm]{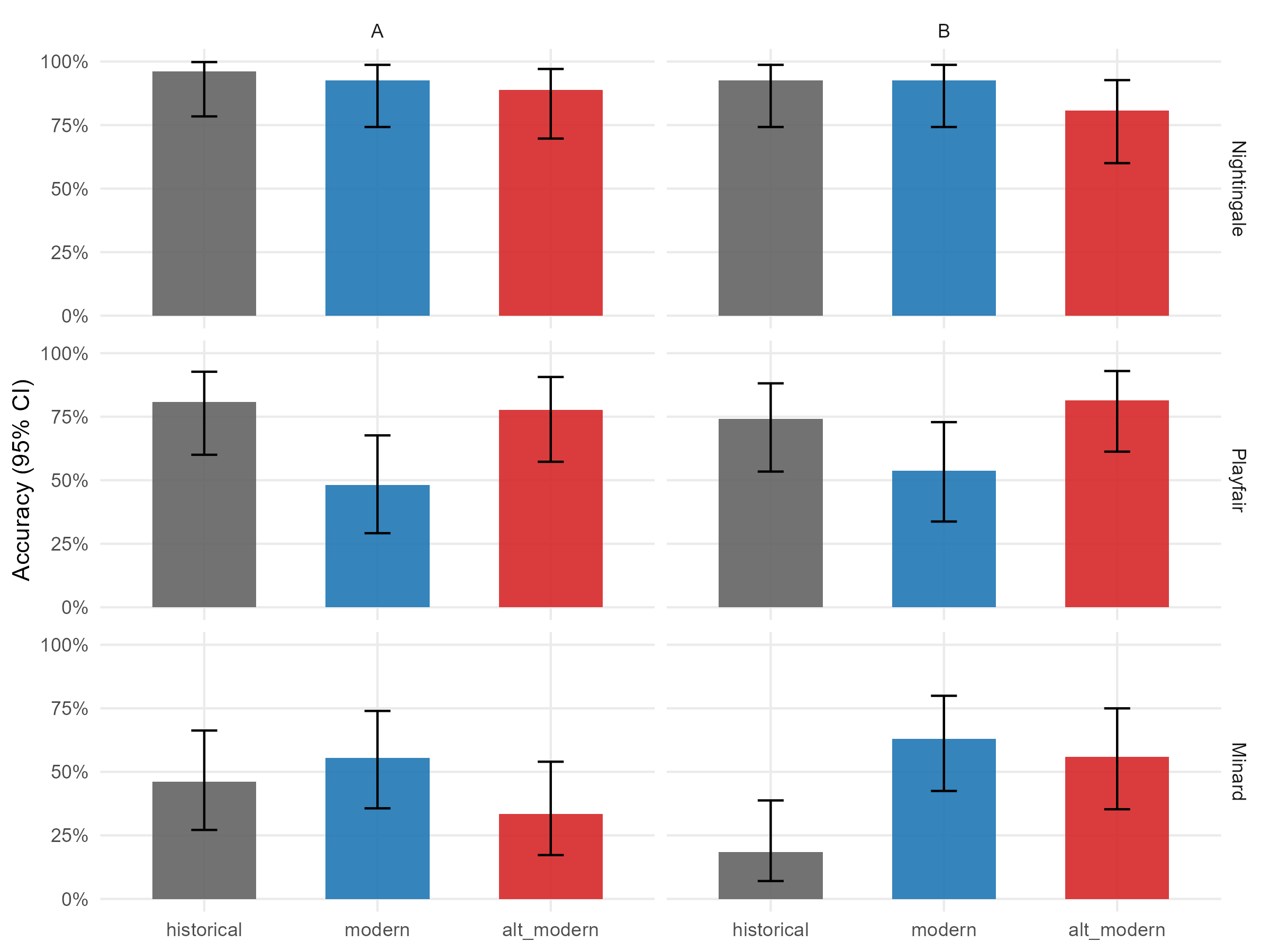}
  \caption{Accuracy by figure and version, split by order (A: N→P→M; B: M→P→N). Historical Minard suffers when first; modern Minard is stable; Playfair dual-axis remains weak regardless of order.}
  \label{fig:order-acc}
\end{figure}

\begin{figure}[H]
  \centering
  \includegraphics[width=\textwidth, height = 8cm]{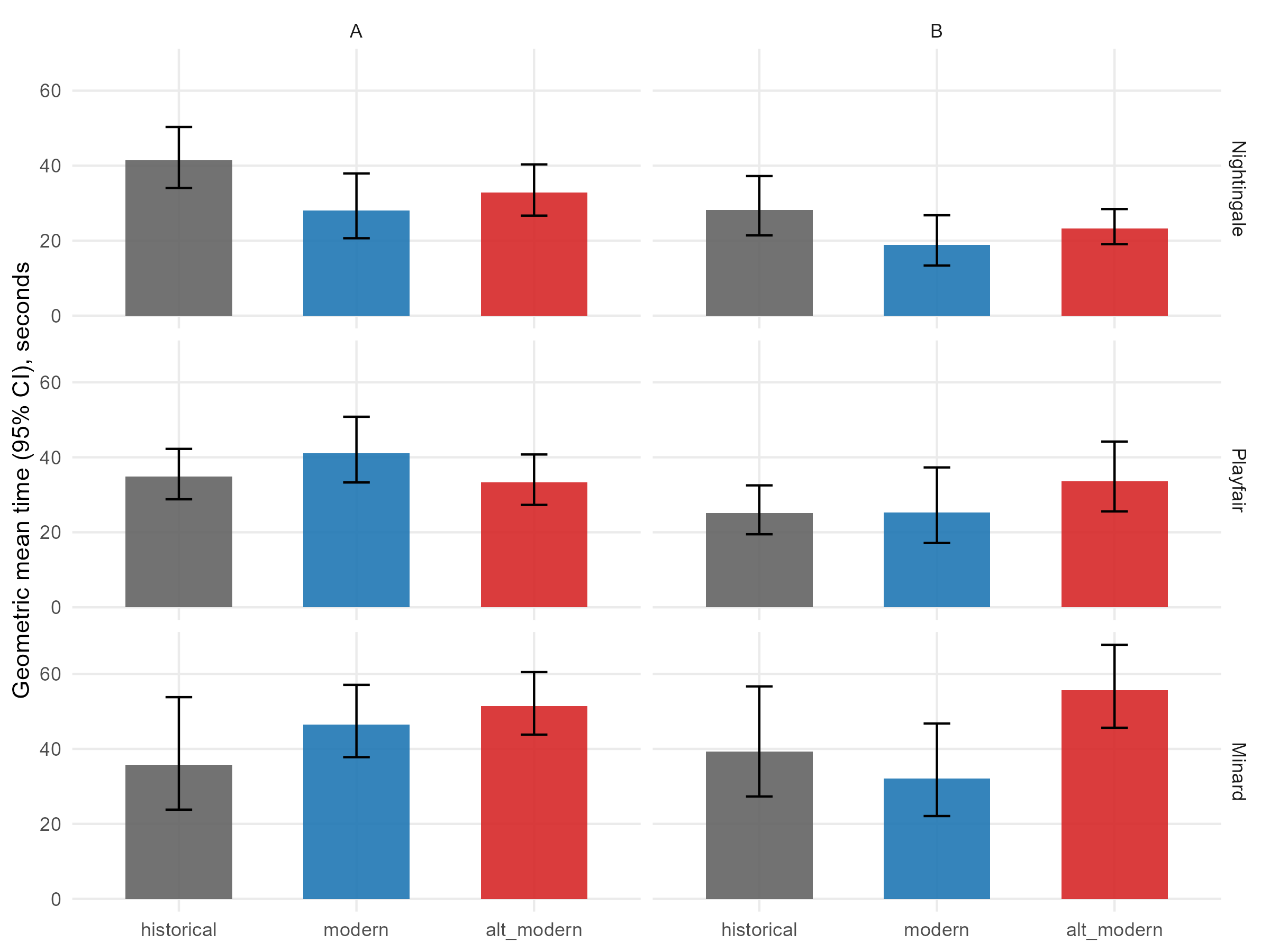}
  \caption{Response time by figure and version, split by order (geometric mean, 95\% CI). Order effects are mild for time, except that Nightingale is faster when it appears last.}
  \label{fig:order-time}
\end{figure}

\printbibliography

\end{document}